\documentclass{ws-ijmpd}
\usepackage[super,compress]{cite}
\usepackage[breaklinks]{hyperref}
\hypersetup{colorlinks,urlcolor=black,citecolor=black,linkcolor=black,filecolor=black}
\usepackage{breakurl}
\usepackage{url}

\usepackage[english]{babel}
\usepackage[utf8]{inputenc}
\usepackage{latexsym}
\usepackage{graphics}
\usepackage{amsmath,amssymb}

\newcommand{\epow}[1]{\mathrm{e}^{#1}}
\newcommand{\oder}[2]{\frac{\partial #1}{\partial #2}}

\begin{document}

\markboth{J. Hlad\'{i}k, C. Posada, and Z. Stuchl\'{\i}k}
{Radial instability of trapping polytropic spheres}

%
\catchline{}{}{}{}{}
%

\title{Radial instability of trapping polytropic spheres}

\author{Jan Hlad\'{i}k}
\address{Institute of Physics and Research Centre of Theoretical Physics and Astrophysics, Faculty of Philosophy and Science, Silesian University in Opava, Bezru\v{c}ovo n\'{a}m. 13, CZ-746\,01 Opava, Czech Republic\\ jan.hladik@fpf.slu.cz}

\author{Camilo Posada}
\address{Institute of Physics and Research Centre of Theoretical Physics and Astrophysics, Faculty of Philosophy and Science, Silesian University in Opava, Bezru\v{c}ovo n\'{a}m. 13, CZ-746\,01 Opava, Czech Republic\\ camilo.posada@fpf.slu.cz}

\author{Zden\v{e}k Stuchl\'{\i}k}
\address{Institute of Physics and Research Centre of Theoretical Physics and Astrophysics, Faculty of Philosophy and Science, Silesian University in Opava, Bezru\v{c}ovo n\'{a}m. 13, CZ-746\,01 Opava, Czech Republic\\ zdenek.stuchlik@fpf.slu.cz}

\maketitle

\begin{history}
\received{Day Month Year}
\revised{Day Month Year}
\end{history}

\begin{abstract}
  We complete the stability study of general relativistic spherically symmetric polytropic perfect fluid spheres, concentrating attention to the newly discovered polytropes containing region of trapped null geodesics. We compare the methods of treating the dynamical stability based on the equation governing infinitesimal radial pulsations of the polytropes and the related Sturm--Liouville eigenvalue equation for the eigenmodes governing the pulsations, to the methods of stability analysis based on the energetic considerations. Both methods are applied to determine the stability of the polytropes governed by the polytropic index $n$ in the whole range $0 < n < 5$, and the relativistic parameter $\sigma$ given by the ratio of the central pressure and energy density, restricted by the causality limit. The critical values of the adiabatic index for stability are determined, together with the critical values of the relativistic parameter $\sigma$. For the dynamical approach we implemented a numerical method which is independent on the choice of the trial function, and compare its results with the standard trial function approach. We found that the energetic and dynamic method give nearly the same critical values of $\sigma$. We found that all the configurations having trapped null geodesics are unstable according to both methods.
\end{abstract}


\ccode{PACS numbers: 04.40.Dg, 95.30.Sf}

\section{Introduction}\label{intro}

The polytropic spherical configurations represent one of the simplest approaches to describe the basic properties of astrophysical objects. They are frequently applied in astrophysics as models of both Newtonian and general relativistic stars. The polytropic spheres are well known representatives of compact objects, as they exemplify dense nuclear matter inside neutron or quark stars. For example, they describe the fluid configurations constituted from non-relativistic ($n = 4/3$) and relativistic ($n = 5/3$) degenerate Fermi gas \cite{Shapiro:1983du,Zeldovich:1971}, considered as basic approximations of neutron star matter.

The polytropic spheres in spacetimes with a non-zero cosmological constant, were studied recently in Ref. \refcite{Stuchlik:2016xiq}  (see also Refs. \cite{Stuchlik:2000xe,Boehmer:2004nu,Boehmer:2003uz}). In this work three important new results were found:
\begin{itemize}
\item[a) ]The polytropes could represent appropriate models not only for compact objects as neutron stars (see also Ref. \refcite{Alvarez-Castillo:2017qki} where realistic equations of state of neutron stars are modelled by several polytropic regions), but also models of extremely extended dark matter halos of galaxies or their clusters, with extension that cannot exceed the so called static radius given by the polytrope mass (or the mass of the halo) \cite{Stuchlik:1999qk,Stuchlik:2009jv,Stuchlik:2011zz}.
\item[b) ] For certain values of the polytropic index $n$, and the relativistic parameter $\sigma$, the polytropes contain an internal region containing trapped null geodesics \cite{Novotny:2017cep,Hod:2018kql}. We call them shortly trapping polytropes.
\item[c) ] In the regions of the trapped null geodesics, the trapping polytropes are unstable against gravitational perturbations, and the gravitational instabilities could lead to gravitational collapse of the regions of trapped null geodesics and creation of supermassive black holes in the central regions of extremely extended polytropes modelling galaxy dark matter halos \cite{Stuchlik:2017qiz}.
\end{itemize}

The issue of the stability of polytropic general-relativistic configurations is not a new one. Tooper \cite{Tooper:1964} investigated some general properties of the polytropic spheres, including their stability for large relativistic parameter $\sigma$, by using energetic considerations. He suggested that the polytrope $n = 3$ could be energetically unstable for $\sigma>0.5$, leading to the possibility of a transition from a higher total energy state towards one of lower energy. Tooper conjectured that these transitions could, in principle, explain the high energy emissions of some radio sources. Tooper's results were corrected and extended by Bludman \cite{Bludman:1973}, who concluded that the polytrope $n = 3$ is unstable for $\sigma > 0$.

Although the energetic considerations might be the most direct approach to study the stability of gas configurations, some caution must be taken when making definite conclusions concerning stability by studying static solutions only. For instance, some thermal properties like the adiabatic index (ratio of specific heats) could play a role in the instability which could not be predicted from purely energy considerations. This ambiguity was addressed by Chandrasekhar \cite{Chandrasekhar:1964zza,Chandrasekhar:1964zz} who developed the theory of infinitesimal, and adiabatic, radial oscillations of general-relativistic polytropes, using a linear analysis of time-dependent perturbations on equilibrium states. The main conclusion of this work is that the Newtonian value of the adiabatic index $\gamma = 4/3$ for stability is increased by relativistic effects.

In this paper we present a detailed study of the stability of the polytropic spheres for the whole range of the parameters governing the structure of the polytropes, using two different approaches, namely, the standard energetic methods introduced in Refs. \refcite{Tooper:1964,Tooper:1965}, which were applied and corrected in Ref. \refcite{Bludman:1973}; and the dynamic method based on the study of radial pulsations introduced in Refs. \refcite{Chandrasekhar:1964zza,Chandrasekhar:1964zz} and discussed in Ref. \refcite{Misner:1974qy}. For simplicity, we assume here a vanishing cosmological constant, $\Lambda = 0$, and concentrate attention to the cases where the polytrope parameters determine the trapping polytropes with the regions of trapped null geodesics that could demonstrate a gravitational instability. The role of the repulsive cosmological constant ($\Lambda > 0$) in the stability problem will be considered in a following separated paper.

In Section~\ref{sec:2} we review the general properties of the relativistic polytropic fluid spheres. In Section~\ref{sec:3} we introduce the dynamical equation for the radial pulsations of spherically symmetric perfect fluid configurations under the assumption of adiabatic processes. The related boundary conditions are specified in a way corresponding to the treatment presented in Ref. \refcite{Misner:1974qy,Kokkotas:2000up}. The Sturm--Liouville equation for the eigenfrequencies of the pulsation eigenmodes is given, and its application to polytropic spheres. In Section~\ref{sec:3.4} we discuss the numerical methods we used to solve the Sturm--Liouville eigenvalue problem. In Section~\ref{sec:3.5} we present our results for the critical adiabatic index for stability, for several polytropes with different index $n$. In Section~\ref{sec:4}, we discuss the energetic approach and its application to the considerations of stability of the polytropic spheres. Concluding remarks are presented in Section~\ref{sec:5}.


\section{General-Relativistic polytropic spheres}\label{sec:2}

Here we consider the models of static polytropic fluid spheres proposed by Tooper \cite{Tooper:1964}, which correspond to the relativistic generalization of the classical Lane--Emden models.\footnote{Polytropic configurations allow the existence of trapped null geodesics, see Ref. \refcite{Novotny:2017cep}.} We also consider the whole variety of polytropic equations of state giving acceptable stellar models. The polytropic spheres are governed by the equation of state
\begin{equation} \label{polytropic}
  p = K\rho^{1+\frac{1}{n}},
\end{equation}
\noindent where $\rho$ is the energy density, $n$ is the polytropic index and $K$ is a constant related to the characteristics of a specific fluid sphere. It is conventional to introduce the relativistic parameter
\begin{equation}\label{sigma}
  \sigma = \frac{p_{\mathrm{c}}}{\rho_{\mathrm{c}} c^2}\, ,
\end{equation}
\noindent where $\rho_\mathrm{c}$ denotes the central energy density and $p_c$ is the central pressure.
The radial profiles of the energy density and pressure of the equilibrium polytropic spheres are given by the relations \cite{Stuchlik:2000xe,Tooper:1964}
\begin{equation}\label{polyeos}
  \rho = \rho_{\mathrm{c}}\theta^n\, ,\qquad p = p_{\mathrm{c}} \theta^{n+1}\, ,
\end{equation}
\noindent where $\theta(x)$ is function of the dimensionless radius
\begin{equation}\label{polyx}
  x \equiv \frac{r}{L}\, ,\qquad L \equiv \left[\frac{\sigma(n + 1)c^2}{4\pi G\rho_{\mathrm{c}}}\right]^{\frac{1}{2}}\, .
\end{equation}
\noindent Here $L$ is the characteristic length scale of the polytropic sphere determined by the polytropic index $n$, the parameter $\sigma$, and the central density $\rho_\mathrm{c}$.

The Einstein equations imply that the function $\theta(x)$ and the `mass' function $v(x)\equiv m(x)/M$ are determined by the set of two differential equations
\begin{gather}
  x^2\frac{\mathrm{d}\theta}{\mathrm{d}x} \frac{1 - 2\sigma(n + 1)v(x)/x}{1+\sigma\theta} + v(x)+ \sigma x\theta \frac{\mathrm{d}v}{\mathrm{d}x} = 0\, ,  \label{radius1}\\
  \frac{\mathrm{d}v}{\mathrm{d}x} = x^2 \theta^n\, . \label{radius2}
\end{gather}
These equations, which can be solved by numerical methods \cite{Stuchlik:2016xiq,Tooper:1964}, give the edge of the polytropic sphere as the first solution $\theta(x_1) = 0$ of Eqs.~(\ref{radius1}) and~(\ref{radius2}).\footnote{Only the case $n=0$, corresponding to a uniform density configuration, can be solved exactly.} The radius and mass of the polytropic sphere are determined by $x_1$ and $v(x_1)$ through the relations
\begin{gather}
  R = L x_1\, , \label{radius}\\
  M = 4\pi L^3 \rho_{\mathrm{c}} v(x_1) = \frac{c^2}{G}L\sigma(n + 1)v(x_1)\, . \label{Mass}
\end{gather}
\noindent The radial metric coefficient of the static configuration then reads
\begin{equation}
  \epow{-2\Psi_0(x)} = 1 - 2\sigma (n + 1)\frac{v(x)}{x}\, ,
\end{equation}
\noindent and the temporal metric coefficient takes the form
\begin{equation}\label{grr}
  \epow{2\Phi_0(x)} = (1 + \sigma\theta)^{-2(n+1)} \left[1 - 2 \sigma\left({n+1}\right)\frac{v(x_1)}{x_1}\right]\, .
\end{equation}
A quantity which will be important in our stability analysis is the mass-radius relation, which can be obtained from \eqref{radius} and \eqref{Mass} to give
\begin{equation}
  \frac{2GM}{c^2R} = \frac{r_{g}}{R} = \frac{2\sigma(n + 1)v(x_1)}{x_1}\, ,
\end{equation}
\noindent where $r_{g} = 2GM/c^2$ is the Schwarzschild radius.


\section{Stability against radial pulsations}\label{sec:3}

We first apply the Chandrasekhar method~\cite{Chandrasekhar:1964zz} for radial stability based on treatment of radial pulsations of the equilibrium configurations determined in previous section.

In the standard Schwarzschild coordinates ($t, r, \theta, \varphi$), the spacetime element of the radially pulsating, spherically symmetric polytrope reads \footnote{We follow the same procedure and notation as Ref. \refcite{Misner:1974qy}.}
\begin{equation}\label{metric1}
  \mathrm{d} s^2 = -\epow{2\Phi}\,\mathrm{d} t^2 + \epow{2\Psi}\,\mathrm{d} r^2 + r^2(\mathrm{d}\theta^2 + \sin^2\theta\,\mathrm{d}\varphi^2)\, ,
\end{equation}
\noindent where the metric coefficients are considered in the general form including the time-dependence
\begin{equation}
  \Psi = \Psi(r,t)\,,\qquad\Phi = \Phi(r,t)\, .
\end{equation}
The matter inside the pulsating configuration is assumed to be a perfect fluid represented by the energy density $\rho(r,t)$ and pressure $p(r,t)$. The unperturbed equilibrium polytrope state, about which the radial pulsations are realised, is given by the functions $\Phi_0(r)$, $\Psi_0(r)$, $\rho_0(r)$, $p_0(r)$.

The pulsating polytropic configuration, considered with perturbed quantities depending on time, is determined by the Einstein equations in the form \cite{Misner:1974qy}
\begin{equation}\label{Gtt}
  \frac{1}{r^2}\left[1-\left(r\epow{-2\Psi}\right)'\right] = \frac{8\pi G}{c^4}\,T_{\;\;t}^{t}\, ,
\end{equation}
\begin{equation}\label{Grr}
  \epow{-2\Psi}\left(\frac{2\Phi'}{r}+\frac{1}{r^2}\right)-\frac{1}{r^2} = \frac{8\pi G}{c^4}\,T_{\;\;r}^{r}\, ,
\end{equation}
\begin{equation}\label{Gtheta}
  -\epow{-2\Phi}\left[\ddot{\Psi}+(\dot{\Psi})^2-\dot{\Phi}\dot{\Psi}\right] + \epow{-2\Psi}\left(\Phi''+ (\Phi')^2 - \Phi'\,\Psi' + \frac{\Phi'-\Psi'}{r}\right)=\frac{8\pi G}{c^4}\,T_{\;\;\theta}^{\theta}\, ,
\end{equation}
\begin{equation}\label{Gtr}
  \epow{-2\Psi}\frac{2\dot{\Psi}}{r}=\frac{8\pi G}{c^4}\,T_{\;\; t}^{r}\, .
\end{equation}
Here the prime (dot) denotes partial derivative with respect to the radial (time) coordinate. For pulsations of a small amplitude, the metric coefficients $\Psi(r,t)$ and $\Phi(r,t)$, and the thermodynamic variables $\rho(r,t)$, $p(r,t)$ and the number density of baryons $n(r,t)$, as measured in fluid's rest frame, can be described by their small Euler variations, defined generally in the form
\begin{equation}
  q(r,t) = q_0(r) + \delta q(r,t)\, ,
\end{equation}
\noindent where the general variables correspond to the quantities $\delta q \equiv(\delta\Phi,\delta\Psi,\delta\rho,\delta p,\delta n)$. The pulsation is represented by the radial displacement $\xi$ of the fluid from the equilibrium position
\begin{equation}
  \xi = \xi(r,t)\, .
\end{equation}
The \emph{Euler perturbations} $\delta q$ are connected to the \emph{Lagrangian perturbations} $\Delta q$ measured by an observer co-moving with the oscillating fluid by the relation
\begin{equation}
  \Delta q (r,t) = q(r + \xi(r, t),t) - q_0 (r) \approx\delta q + q_0'\xi\, .
\end{equation}

The pulsation dynamics is determined by the Einstein equations together with the energy-momentum conservation, baryon conservation, and the thermodynamic laws. All the relevant equations must be \emph{linearized} relative to the displacement from the static equilibrium configuration. We have to obtain the dynamic equation for evolution of the fluid displacement $\xi(t,r)$, and a set of \emph{initial-value equations}, expressing the perturbation functions $\delta\Phi,\delta\Psi,\delta\rho,\delta p,\delta n$ in terms of the displacement function $\xi(t,r)$.

\subsection{Energy density, pressure, and metric perturbations}\label{sec:3.1}

No nuclear reactions are assumed during small radial perturbations, thus the dynamics of the energy density and the pressure perturbations is governed by the baryon conservation law. Following Refs. \refcite{Chandrasekhar:1964zz,Misner:1974qy}, we express the velocity of the fluid element in terms of the displacement evolution and define
\begin{equation}
  \frac{u^r}{u^t} = \frac{\partial\xi}{\partial t}\equiv\dot{\xi}\, .
\end{equation}
\noindent The conservation of the number of baryons of the fluid implies
\begin{equation}
  \left(nu^{\mu}\right)_{;\mu} = 0\, .
\end{equation}
\noindent In terms of the Lagrangian perturbation this equation takes the form
\begin{equation}
  \oder{(\Delta n)}{\tau} = -n \left(u^\mu{}_{;\mu}\right)\, .
\end{equation}
\noindent The linearized expressions for the 4-velocity components are given by
\begin{equation}
  u^t = \epow{-\Phi_0}\left(1 - \delta \Phi \right)\, ,\qquad  u^r = \dot{\xi}\epow{-\Phi_0}\, ,
\end{equation}
\noindent which imply the equation
\begin{equation}
\Delta n = -n_{0}\left[\frac{1}{r^2 \epow{\Psi_0}}\left(r^2\epow{\Psi_0}\xi\right)'
    +\delta\Psi\right]\,.
\end{equation}
Focusing attention to the adiabatic pulsations, the Lagrange variables in the number density and the pressure are connected through the adiabatic index $\gamma$ determined by the relation
\begin{equation}\label{gamma}
  \gamma\equiv\left(\frac{\partial \ln p}{\partial \ln n}\right)_{S} = \left(p\,\frac{\partial n}{\partial p}\right)^{-1} \left[n-(\rho+p)\frac{\partial n}{\partial\rho}\right]\, ,
\end{equation}
\noindent which governs the linear perturbations of pressure inside the star. It is worthwhile to remark that this $\gamma$ is not necessarily the same as the adiabatic index associated to the EOS. To compute the $\gamma$ in  \eqref{gamma} one needs not only information about the EOS, but also one needs to know the perturbed EOS under the assumption of constant entropy $S$ and thermodynamic equilibrium of its constituents \cite{Shapiro:1983du,Gondek:1997fd}.

In terms of $\gamma$, the initial value equation for the pressure perturbation reads
\begin{equation}
  \delta p = - \gamma p_0 \left[\frac{\left(r^2\epow{\Psi_0}\xi\right)'}{r^2 \epow{\Psi_0}} + \delta \Psi\right] - \xi p_0'\, .
\end{equation}
Applying projection of the energy-momentum conservation law $T^{\mu\nu}{}_{;\nu} = 0$ onto the 4-velocity $u^{\mu}$ we arrive to the local energy conservation law
\begin{equation}
  \Delta\rho = \frac{\rho_0 + p_0}{n_0}\,\Delta n
\end{equation}
\noindent which allows to write the initial-value equation for the Lagrangian perturbation of the energy density $\delta\rho$ in the form
\begin{equation}
  \delta\rho = -(\rho_0+p_0)\left[\frac{\left(r^2\epow{\Psi_0}\xi\right)'}{r^2 \epow{\Psi_0}} + \delta \Psi\right] - \xi\rho_0'\, .
\end{equation}
In order to obtain the initial-value problem of the metric perturbations, we need the perturbed stress energy tensor components which, in linearized form, are given by
\begin{equation}
  T_{rt} = -(\rho_0 + p_0)\,\epow{\Psi_0 - \Phi_0}\dot{\xi}\, ,\qquad T_{rr} = p_0 + \delta p \, .
\end{equation}
Linearization of \,(\ref{Gtr}), $G_{tr} = (8\pi G/c^4) T_{tr}$, implies the initial-value equation for $\delta\Psi$ in the form
\begin{equation}
  \delta\Psi = -4\pi\left(\rho_0 + p_0\right)r\epow{2\Psi_0\xi} = - \left(\Psi_0' + \Phi_0' \right)\xi\, ,
\end{equation}
\noindent and \,(\ref{Grr}), $G_{rr}=(8\pi G/c^4)T_{rr}$, implies the initial-value equation for $\delta\Phi$ in the form 
\begin{equation}
  \delta\Phi' = -\frac{\gamma}{r}\left(4\pi p_0 \right) \epow{2\Psi_0 + \Phi_0} \left(r^2\epow{-\Phi_0}\xi\right)' +
  \frac{4\pi G}{c^4}\left[p_0'r-\left(\rho_0+p_0\right)\right]\epow{2\Psi_0}\xi\, .
\end{equation}

\subsection{The pulsation equation}\label{sec:3.2}

The dynamics of the fluid small displacements $\xi(t,r)$ is determined by the Euler equation for the four-acceleration $a_\mu$ of the fluid elements. It can be obtained from the projection of the energy-momentum conservation law $T^{\mu\nu}{}_{;\nu} = 0 $ onto the plane orthogonal to $u^{\mu}$
\begin{equation}
  (\rho + p)a_{\mu}=-p_{,\mu}-u_{\mu}u^{\nu}p_{,\nu}\, .      \label{rce_24}
\end{equation}
In the linearized form (\ref{rce_24}), the four-acceleration has only one nonzero component
\begin{equation}
  a_r = \Phi_0' + \delta\Phi' + \epow{2\left(\Psi_0 - \Phi_0\right)}\ddot{\xi}\, .
\end{equation}
\noindent Introducing a ‘renormalized displacement function' $\zeta$ \cite{Misner:1974qy}
\begin{equation}
  \zeta\equiv r^2 \epow{-\Phi_0}\xi\, ,
\end{equation}
and applying the initial-value equations, we obtain the dynamic pulsation equation
\begin{equation}\label{dynamiceq}
  W\ddot{\zeta} = \left(P\zeta'\right)' + Q\zeta\, ,
\end{equation}
where the functions $W(r)$, $P(r)$ and $Q(r)$ are given by
\begin{equation}\label{Wr}
  W \equiv (\rho_0 + p_0)\frac{1}{r^2}\,\epow{3\Psi_0 + \Phi_0}\, ,
\end{equation}
\begin{equation}\label{Pr}
  P \equiv \gamma p_0 \frac{1}{r^2}\,\epow{\Psi_0 + 3\Phi_0}\, ,
\end{equation}
\begin{equation}\label{Qr}
  Q \equiv \epow{\Psi_0 + 3\Phi_0}\left[\frac{(p_0')^2}{\rho_0 + p_0}\frac{1}{r^2} - \frac{4p_0'}{r^3} (\rho_0 + p_0)\left(\frac{8\pi G}{c^4}p_0\right)\frac{\epow{2\Psi_0}}{r^2}\right]\, .
\end{equation}
The boundary conditions must guarantee that the displacement function is not resulting in a divergent energy density and pressure perturbations at the centre of the sphere. On the other hand, the variations of the pressure must satisfy the condition $p(R) = 0$ at the surface of the configuration, therefore we have
\begin{equation}\label{bco1}
  \frac{\xi}{r}\quad\text{is finite, or zero, as}\quad r \rightarrow 0\, ,
\end{equation}
\begin{equation}\label{bco2}
  \Delta p = -\gamma p_0\frac{\epow{\Phi_0}}{r^2}\left(r^2\epow{-\Phi_0}\xi\right)\rightarrow 0\quad \text{as}\quad r\rightarrow R\, .
\end{equation}

For the linear dynamical stability analysis we follow the standard assumption of sinusoidal time-dependence \cite{Chandrasekhar:1964zz}
\begin{equation}
  \zeta(r,t)=\zeta(r)\epow{-\mathrm{i}\omega t}\, ,
\end{equation}
thus the dynamic equation \eqref{dynamiceq} takes the familiar form of the Sturm--Liouville equation \cite{Misner:1974qy}
\begin{equation}
  \left(P\zeta'\right)' + (Q+\omega^2W)\zeta = 0\, ,             \label{eSL}
\end{equation}
and the boundary conditions \eqref{bco1} and \eqref{bco2} which read
\begin{alignat}{2}
  &\frac{\zeta}{r^3}&\quad\text{is finite, or zero, as}&\qquad           \label{bc1}
  r \rightarrow 0\,,\\
  &\gamma p_0 \frac{\epow{\Phi_0}}{r^2}\zeta' \rightarrow 0            \label{bc2}
  &\text{as}&\qquad r\rightarrow R\, .
\end{alignat}
The Sturm--Liouville equation \eqref{eSL} together with the boundary conditions \eqref{bc1} and \eqref{bc2}, determine the pulsation eigenfunctions $\zeta_i(r)$ and eigenvalues $\omega_i$, where $i = 1,2,\ldots$ The eigenvalue Sturm--Liouville (SL) problem can be expressed in the variational form of Ref. \refcite{Misner:1974qy}, because the extremal values of
\begin{equation}\label{e36}
\omega^2 = \frac{\int_0^R\left(P\zeta'^2-Q\zeta^2\right)\,\mathrm{d}r}{\int_0^R W\zeta^2\,\mathrm{d}r}
\end{equation}
\noindent determine the eigenfrequencies $\omega_i$. The absolute minimum value of \,\eqref{e36} corresponds to the squared frequency of the fundamental mode of the radial pulsations. If $\omega^2$ is positive (negative) the configuration is stable (unstable) against radial oscillations. Moreover, if the fundamental mode is stable ($\omega_0^2 > 0$), all higher radial modes will also be stable. For this reason, a sufficient condition for the dynamical instability is the vanishing of the right hand side of \,\eqref{e36} for certain trial function satisfying the boundary conditions \cite{Chandrasekhar:1964zz,Misner:1974qy}.

The Sturm--Liouville equation can be used to determine the dynamical instability of spherical configurations of perfect fluid with any equation of state. In the case of the uniform density polytropes with index $n=0$, the critical adiabatic index $\gamma_{\mathrm{c}}$, given by the condition $\omega=0$, can be determined by direct integration of the Sturm--Liouville equation. For configurations where the relativistic effects are small $(GM/(c^2R)\ll\,1)$, Chandrasekhar \cite{Chandrasekhar:1964zz} showed that the condition for radial stability of the uniform polytropic spheres is modified from its Newtonian value by

\begin{equation}
  \gamma > \gamma_{\mathrm{c}} \equiv \frac{4}{3} + \frac{19}{42}\frac{r_{\mathrm{g}}}{R}\, ,
\end{equation}
where $R$ corresponds to the radius of the star. This result implies that in Einstein's theory, spherical configurations are more easily destabilised under radial perturbations, compared to the Newtonian theory.

\subsection{Sturm--Liouville equation for polytropic spheres}\label{sec:3.3}

Using the relevant expressions for relativistic polytropes, discussed in Section~\ref{sec:2}, we arrive to the Sturm--Liouville equation for dynamical stability of the polytropic spheres with respect to radial pulsations
\begin{multline}\label{pulsationx}
  \omega^2L^2\int_0^{x_1}\epow{3\Psi_0 + \Phi_0}\theta^n(1 + \sigma\theta)\zeta^2\,\frac{\mathrm{d} x}{x^2} =
  \sigma\int_0^{x_1}\gamma \epow{\Psi_0 + 3\Phi_0}\theta^{n+1}\left(\oder{\zeta}{x}\right)^2\frac{\mathrm{d} x}{x^2}\\
  -\sigma(n + 1)\int_0^{x_1}\epow{\Psi_0 + 3\Phi_0}\left\{\theta^n\left(\oder{\theta}{x}\right)
  \frac{4}{x}\left[\frac{\sigma(n + 1)x}{4(1 + \sigma\theta)}\oder{\theta}{x}-1\right]-2\sigma(1+\sigma\theta)\theta^{2n + 1}\epow{2\Psi_0}\right\} \zeta^2\,\frac{\mathrm{d} x}{x^2}\, .
\end{multline}
For the polytropic spheres the adiabatic index $\gamma$ is given by
\begin{equation}\label{gamma_poly}
  \gamma = \left(1 + \frac{1}{n}\right)(1 + \sigma\theta)\, ,
\end{equation}
\noindent which is a function of the radial coordinate. Chandrasekhar \cite{Chandrasekhar:1964zz} assumed $\gamma$ to be a constant, which is equivalent to consider $\gamma$ in \eqref{pulsationx} as an ‘effective' adiabatic index \cite{Merafina:1989}
\begin{equation}\label{effective}
  \langle \gamma \rangle = \frac{\displaystyle \int\limits_0^{x_1}\gamma\,\epow{\Psi_0 + 3\Phi_0} \frac{\theta^{n + 1}}{x^2}\left(\oder{\zeta}{x}\right)^2\mathrm{d}x}{\displaystyle \int\limits_0^{x_1}\epow{\Psi_0 + 3\Phi_0}\frac{\theta^{n+1}}{x^2}\left(\oder{\zeta}{x}\right)^2\mathrm{d} x}\, .
\end{equation}
\noindent Thus, the condition for stability implies that
\begin{equation}
  \langle \gamma \rangle > \gamma_\mathrm{c}\, ,
\end{equation}
\noindent where $\gamma_\mathrm{c}$ is considered to be the effective critical value of the adiabatic index for the marginally stable case $\omega^2 = 0$. The relation of the radial derivatives of $p$ and $\Phi$ is transferred into the form
\begin{equation}
  \oder{\Phi}{x} = -\frac{(n + 1)\sigma}{1 + \sigma\theta}\oder{\theta}{x}\, .
\end{equation}
In terms of the variables introduced in Eqs. \eqref{sigma}--\eqref{polyx}, the Sturm--Liouville equation \eqref{eSL} takes the form
\begin{equation}\label{eSLx}
  \frac{\mathrm{d}}{dx}\left[P(x)\frac{d\zeta}{dx}\right] + L^2\left[Q(x) + \omega^2\,W(x)\right]\zeta(x)=0\, ,
\end{equation}
\noindent where the functions $W$, $P$, and $Q$ given by Eqs.~\eqref{Wr}, \eqref{Pr} and \eqref{Qr} are now
\begin{equation}\label{Wx}
  W(x) = \frac{\rho_\mathrm{c}\theta^{n}(1 + \sigma\theta)}{L^2 x^2}\epow{3\Psi_0 + \Phi_0}\, ,
\end{equation}
\begin{equation}\label{Px}
  P(x) = \frac{\langle\gamma\rangle\sigma\rho_\mathrm{c}\theta^{n + 1}}{L^2 x^2}\epow{\Psi_0 + 3\Phi_0}\, ,
\end{equation}
\begin{equation}\label{Qx}
  Q(x) = \frac{\sigma\rho_\mathrm{c}(n + 1)\theta^{n}\epow{\Psi_0 + 3\Phi_0}}{L^4 x^2}\left[\frac{\sigma(n + 1)}{(1 + \sigma\theta)}\left(\frac{\mathrm{d}\theta}{\mathrm{d}x}\right)^2 - \frac{4}{x}\frac{\mathrm{d}\theta}{\mathrm{d}x}-2\sigma(1+\sigma\theta)\theta^{n+1}\epow{2\Psi_0}\right]\, .
\end{equation}
In the next section we will discuss methods to solve the eigenvalue problem \eqref{eSLx}, subject to the boundary conditions \eqref{bc1} and \eqref{bc2}, for polytropic spheres.

\subsection{Numerical Methods}\label{sec:3.4}

Clearly, for general polytropes, the critical value of the adiabatic index related to the dynamical stability can be determined by numerical integration only.

Several methods to solve the eigenvalue problem \eqref{pulsationx} have been described in the literature (see e.g. Ref. ~\refcite{Bardeen:1966} and references therein). Following Ref. \refcite{Chandrasekhar:1964zz}, we computed the critical values of the adiabatic index $\gamma_\mathrm{c}$, for the onset of instability, by integrating numerically \eqref{eSLx} in the case $\omega^2 = 0$. For that purpose we followed two different methods: the shooting method and trial functions.

In the shooting method one integrates \eqref{eSLx} from the center up to the surface of the star with some trial value of $\gamma$. The value for which the solution satisfies (within a prescribed error) the boundary conditions \eqref{bc1} and \eqref{bc2} corresponds to the critical adiabatic index $\gamma_\mathrm{c}$.

In order to apply the shooting method to \eqref{eSLx}, it is convenient to transform it to a set of two ordinary differential equations. We follow the convention used in Ref. \refcite{Kokkotas:2000up} where
\eqref{eSLx} can be split in the following form
\begin{gather}
  \frac{\mathrm{d}\zeta}{\mathrm{d}x} = \frac{\eta}{P(x)}\, ,  \label{ode1}\\
  \frac{\mathrm{d}\eta}{\mathrm{d}x} = -L^2\,\left[\omega^2\,W(x) + Q(x)\right]\zeta\, ,   \label{ode2}
\end{gather}
\noindent which satisfy the following behaviour near the origin
\begin{gather}
  \zeta(x) = \frac{\eta_{0}}{3P(0)}x^3 + \mathcal{O}(x^5)\, ,  \label{origin1}\\
  \eta(r) = \eta_{0}\, , \label{origin2}
\end{gather}
\noindent where $\eta_{0}$ is an arbitrary constant, which we choose to be unity \cite{Kokkotas:2000up}.

The second method is based on using trial functions to integrate \eqref{pulsationx}. Following Ref. \refcite{Chandrasekhar:1964zz} we chose the following functions
\begin{equation}
  \xi_1 = x \epow{\Phi_0/2}\, ,\qquad \xi_2 = x\, ,     \label{Chandratrial}
\end{equation}
yielding
\begin{equation}
  \zeta_1 = x^3\epow{-\Phi_0/2}\, ,\qquad \zeta_2 = x^3 \epow{-\Phi_0}\, .
\end{equation}
We realize the detailed study of the stability for the whole range of the polytropes respecting the condition of causality due to the restriction on the relativistic parameter~\cite{Tooper:1964}
\begin{equation}\label{causal}
  \sigma < \sigma_\mathrm{causal} \equiv \frac{n}{n + 1}\, .
\end{equation}
Note however that this restriction applies only for isentropic configurations ($S = \mathrm{constant}$) \cite{Horedt:2004}. Moreover, condition \eqref{causal} is obtained from the relation
\begin{equation}\label{vsc}
  \textit{\texttt{v}}_\mathrm{sc} = c \left(\frac{n + 1}{n}\sigma\right)^{1/2}\, ,
\end{equation}
\noindent which corresponds to the speed of sound at the center of the star. Thus it might seem that  \eqref{vsc} implies the restriction \eqref{causal}. However, \eqref{vsc} gives the phase velocity which is not the same as the group velocity, therefore condition \eqref{causal} might not be definitive.

Applying the methods described above, we have computed critical values of the adiabatic index $\gamma$, for polytropes with characteristic values of the polytropic index $n$. Using these results for $\gamma_\mathrm{c}$, we also found constraints on the relativistic parameter $\sigma$ in order to construct stable configurations. We present our results in the next section.

\subsection{Results of stability given by Chandrasekhar's method} \label{sec:3.5}
\begin{figure}
  \centering
  \includegraphics[width=0.49\textwidth,keepaspectratio=true]{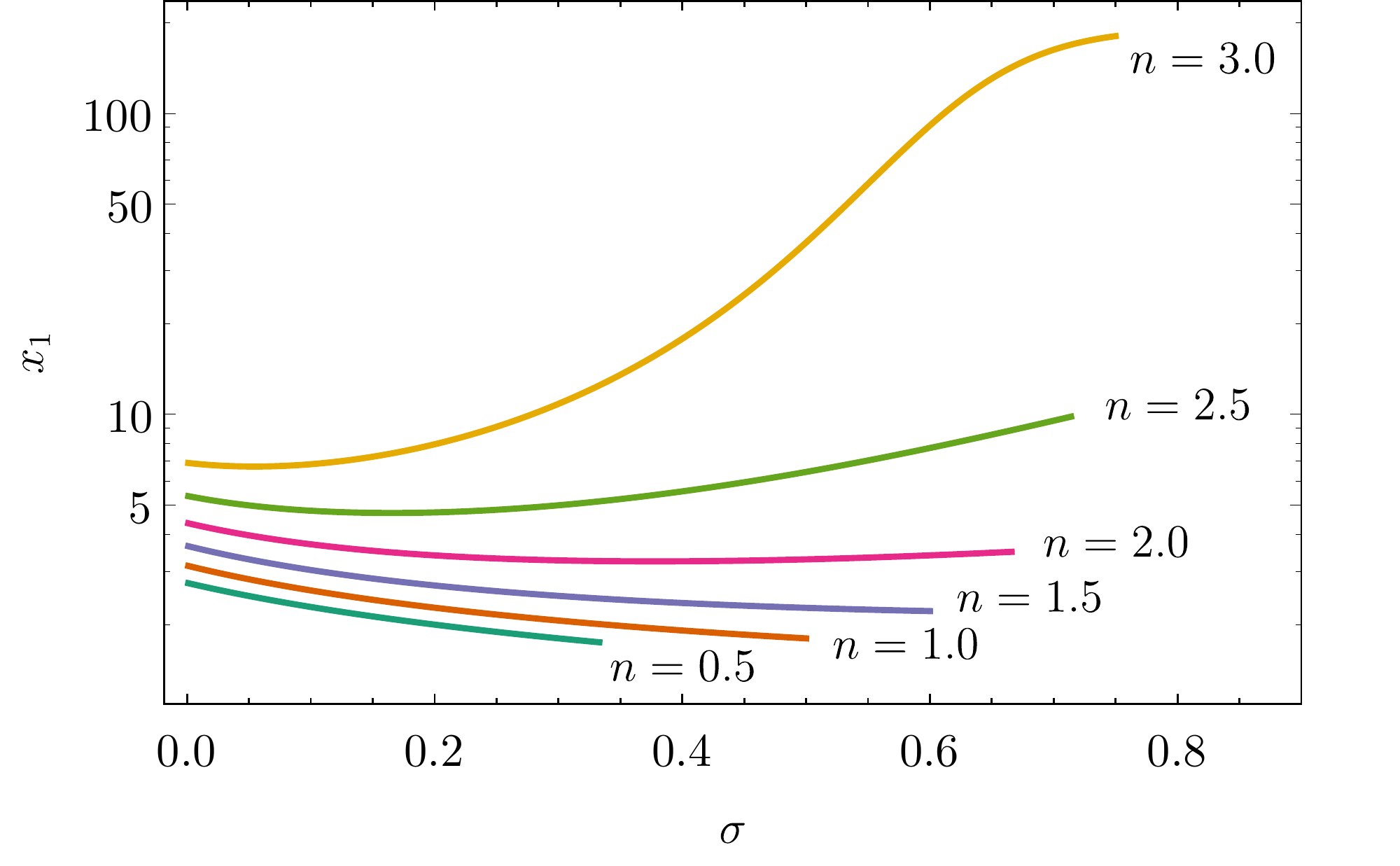}\hfill
  \includegraphics[width=0.49\textwidth,keepaspectratio=true]{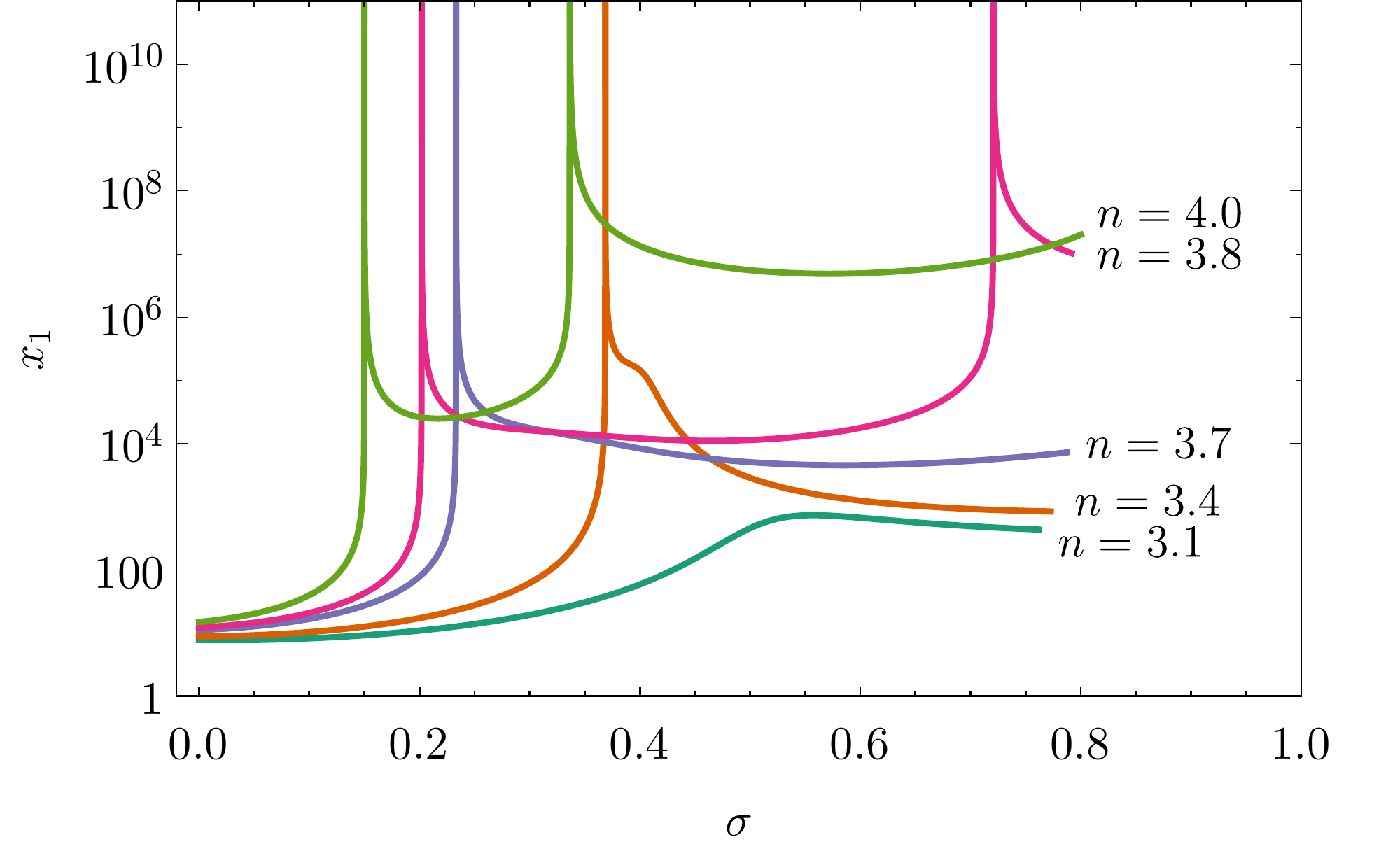}
  \caption{Radius $x_{1}$ as a function of the relativistic parameter $\sigma$. Left panel: polytropes in the range $0 < n \leq 3$. Right panel: polytropes in the range $3 < n \leq 4$.\label{fig1}}
\end{figure}

As a first step in our analysis we solved the structure equations~\eqref{radius1} and~\eqref{radius2}, for a whole family of polytropic spheres~\cite{Stuchlik:2016xiq}, and we briefly summarize the results of this study related to the extension of the polytropes in dependence on the parameters $n$, $\sigma$. In Fig.~\ref{fig1} we show typical dimensionless radii $x_1$ for typical polytropes in the range $0 < n < 4$, as a function of the relativistic parameter $\sigma$. Notice that the polytrope radii dependencies $x_1(\sigma, n)$ are separated into two groups, depending on the polytropic index $n$. For the range $0 < n \leq 3$ the profile is decreasing with $\sigma$ increasing for $n < 1.5$, and a minimum of the profile occurs for $1.5 \leq n \leq 3$. As shown in Ref. \refcite{Nilsson:2000zg} for $n > 3.339$ the profile $x_1(\sigma, n)$ diverges for some critical value(s) of $\sigma = \sigma_\mathrm{\inf}(n)$ and the polytropic spheres do not exist for $n > 5$.

\begin{table}[thb]
  \tbl{Critical adiabatic index $\gamma_{\rm c}$ for radial stability of a relativistic polytropic sphere $(n = 3)$ for some values of the relativistic parameter $\sigma$. For values of $\sigma < 0.1$, we obtained good agreement with the values reported by Chandrasekhar \cite{Chandrasekhar:1964zz}.}
  {\begin{tabular}{ccccc} \toprule
  & & & \multicolumn{2}{c} {Chandrasekhar \cite{Chandrasekhar:1964zz}}\\
  \cline{4-5}
  $\sigma$ & $x_{1}$ & $\gamma_\mathrm{c}$ & $\gamma_\mathrm{c}(\xi_{1})$ & $\gamma_\mathrm{c}(\xi_{2})$ \\ \colrule
  0.015 & 6.803367 & 1.373303 & 1.3732 & 1.3732 \\
  0.040 & 6.718708 & 1.441973 & 1.4411 & 1.4411\\
  0.100 & 6.825877 & 1.616258 & 1.6088 & 1.6088\\
  0.200 & 7.950708 & 1.932184 & 1.8894 & 1.9144\\ \botrule
\end{tabular}\label{table1}}
\end{table}

\begin{figure}
\begin{minipage}[b]{.49\textwidth}
  \centering
    \includegraphics[width=\linewidth,keepaspectratio=true]{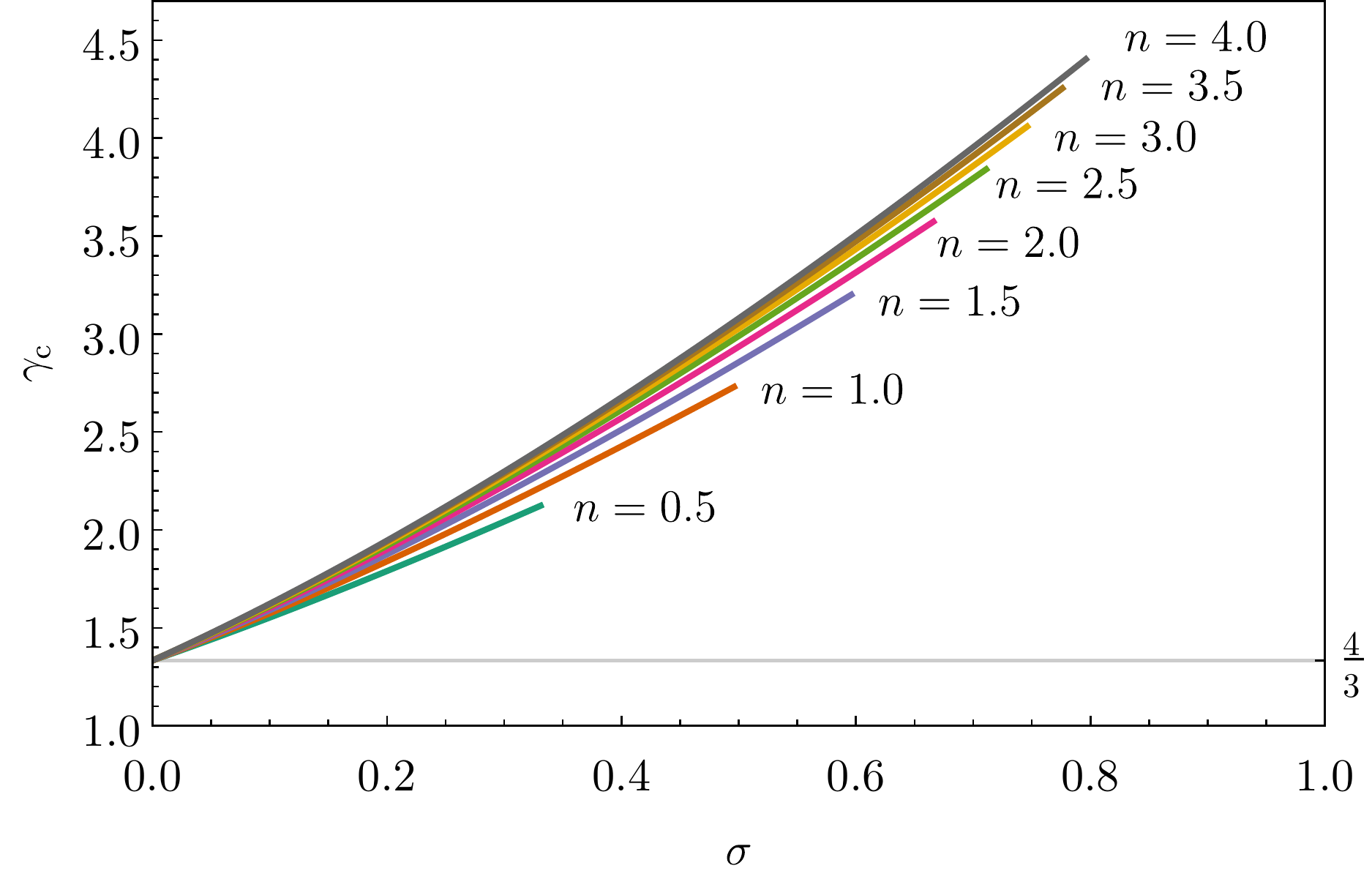}
    \caption{Critical adiabatic index $\gamma_\mathrm{c}$ for stability, as a function of the relativistic parameter $\sigma$ for several values of the polytropic index $n$.\label{fig2}}
\end{minipage}\hfill%
\begin{minipage}[b]{.49\textwidth}
    \centering
    \includegraphics[width=\linewidth,keepaspectratio=true]{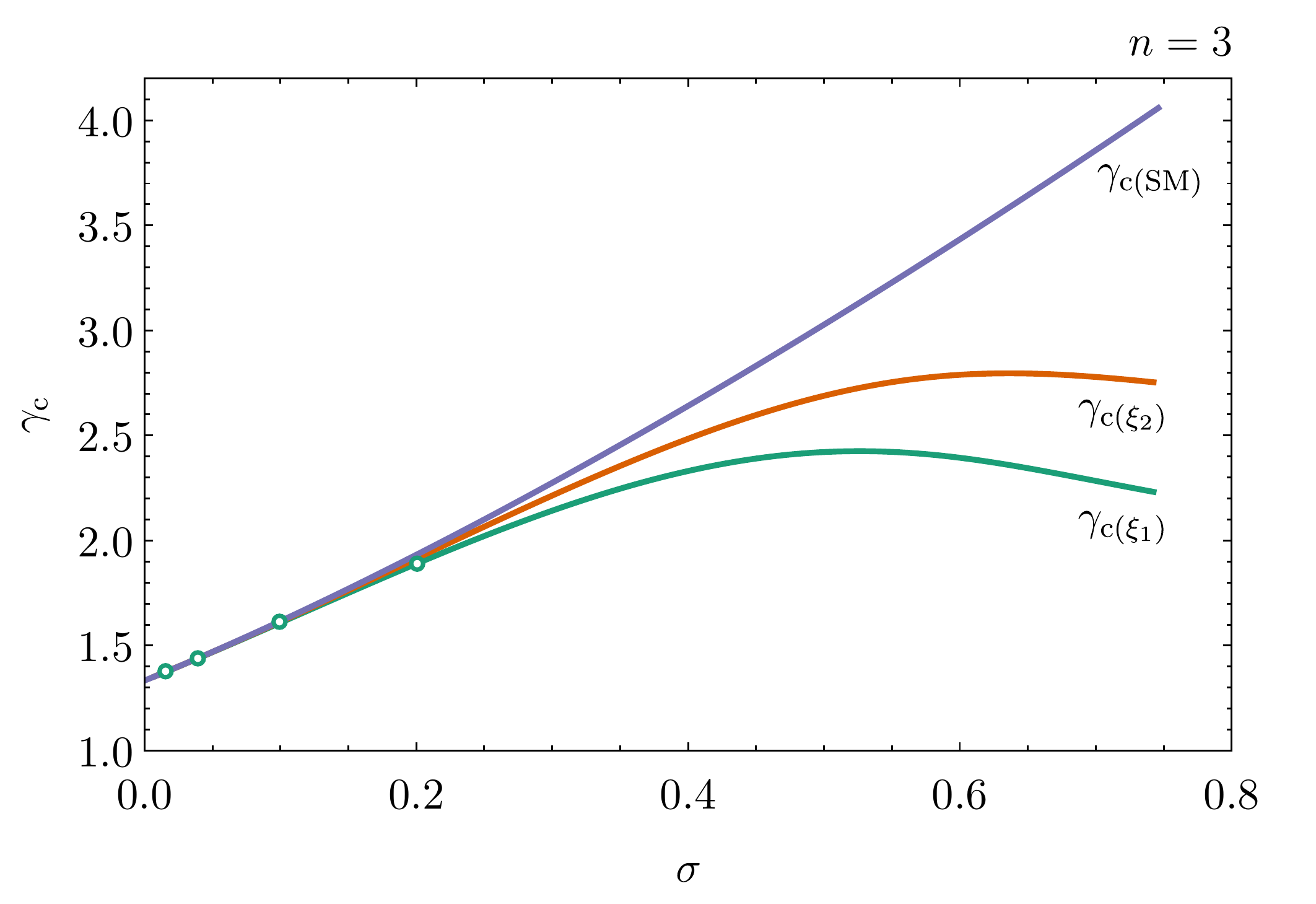}
  \caption{Critical adiabatic index $\gamma_\mathrm{c}$ for a polytrope with index $n = 3$. The dots indicate the results reported in \cite{Chandrasekhar:1964zz} for the trial function $\xi_1$.}\label{fig3}
\end{minipage}
\end{figure}

We computed, via the shooting method (SM), the critical adiabatic index $\gamma_\mathrm{c}$ for different polytropic spheres. We present our results in Fig.~\ref{fig2}. For the configuration $n = 3$, we obtained good agreement with the results reported by Chandrasekhar \cite{Chandrasekhar:1964zz} which were computed using two different trial functions $\xi_{1}$ and $\xi_{2}$ (for comparison see Table~\ref{table1} and Fig.~\ref{fig3}). Note that our result for $\gamma_\mathrm{c}$ obtained in the case of $\sigma = 0.2$, is in better agreement with the result of Chandrasekhar, if determined by using the trial function $\xi_{2}$.

In Fig.~\ref{fig3} we also plot the values of $\gamma_\mathrm{c}$ as computed using the trial functions $\xi_{1}$ and $\xi_{2}$, for the polytrope $n = 3$. Differences are appreciable for higher values of $\sigma$. Notice that our method is independent of the trial functions, moreover we were able to find the optimal eigenfunction corresponding to the fundamental, or marginally stable, mode.

Note that in the nonrelativistic limit when $\sigma \to 0$, the critical adiabatic index approaches the Newtonian value $4/3$. In general, in the regime of low $\sigma$, dynamical stability requires $\gamma > 4/3 + \delta$, where $\delta$ is a quantity proportional to $\sigma$. Thus, in this regime the post-Newtonian approximation is valid \cite{Chandrasekhar:1964zz}. For higher values of $\sigma$, the critical $\gamma$ depends quadratically on the relativistic parameter $\sigma$.

\begin{figure}
  \centering
  \includegraphics[width=0.485\textwidth,keepaspectratio=true]{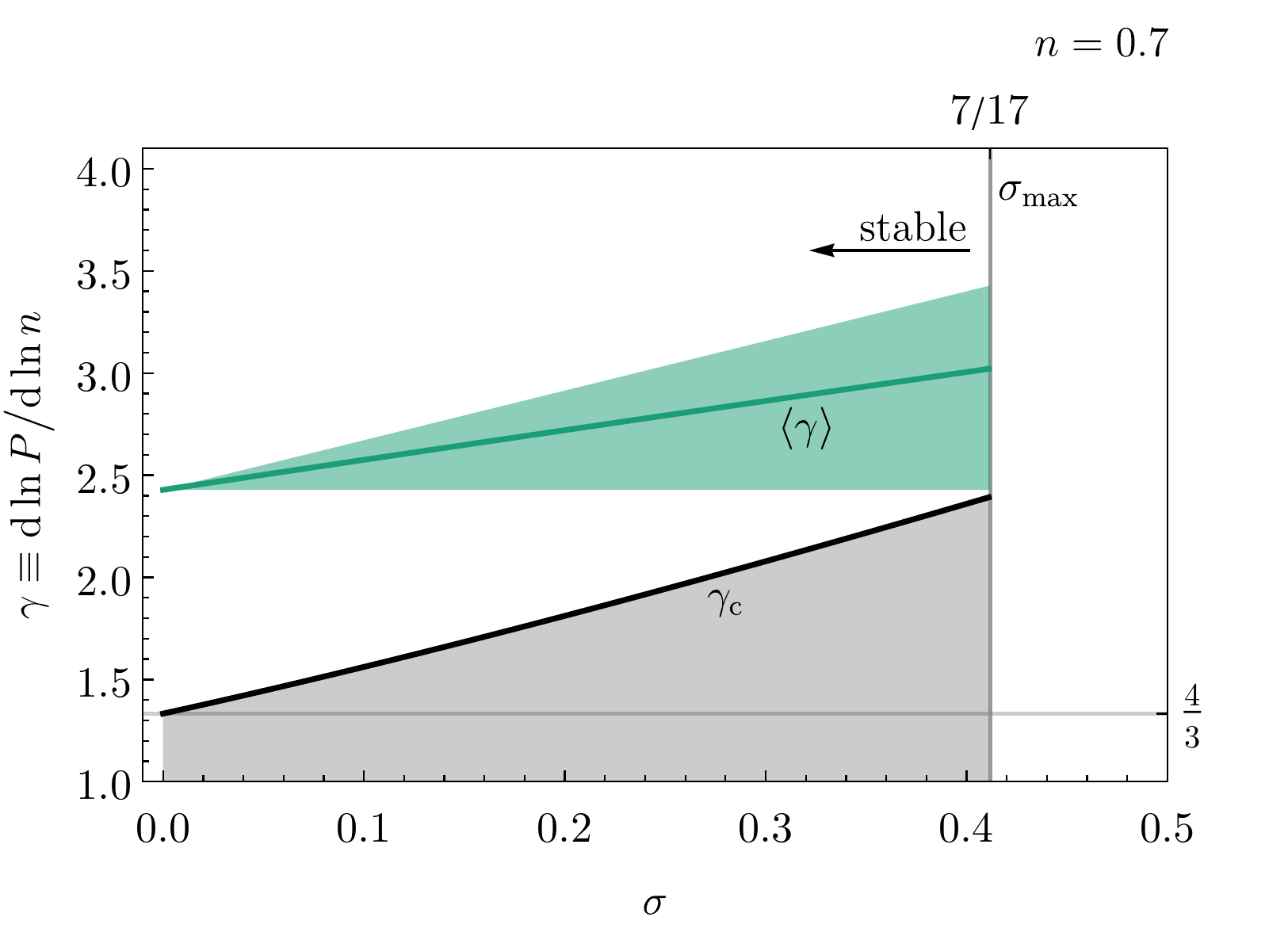} \hfill \includegraphics[width=0.485\textwidth,keepaspectratio=true]{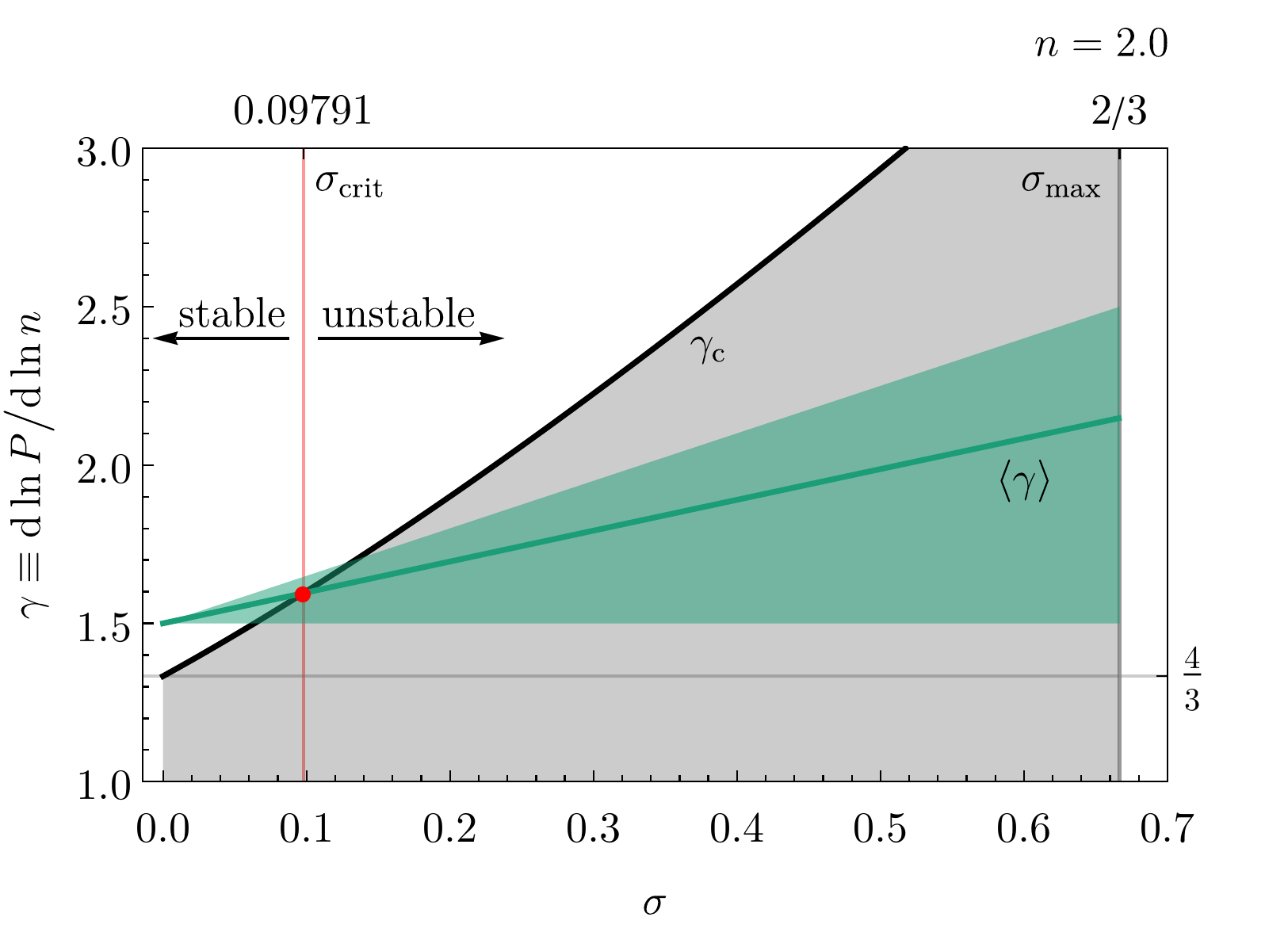}\\
  \includegraphics[width=0.485\textwidth,keepaspectratio=true]{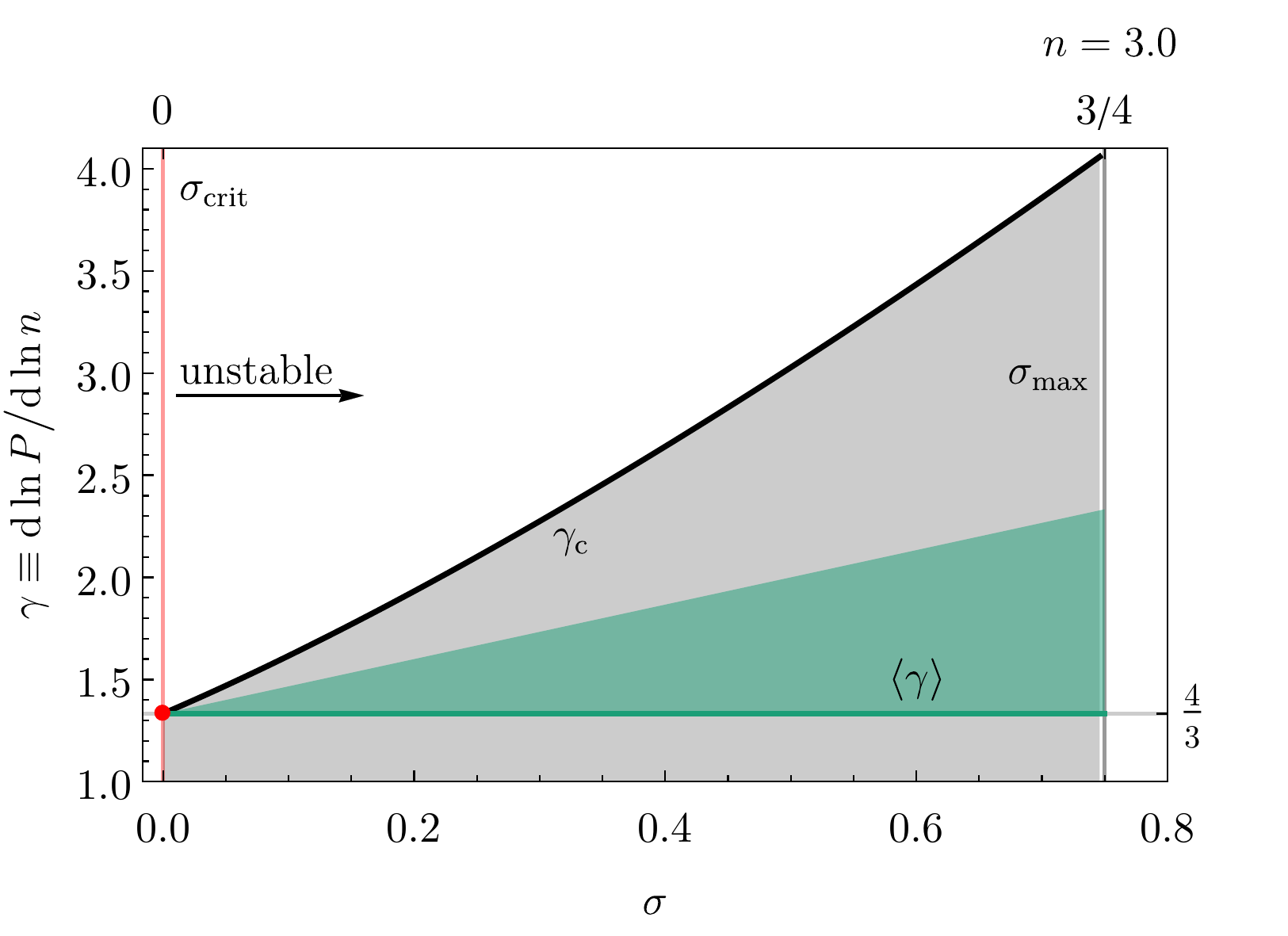} \hfill \includegraphics[width=0.485\textwidth,keepaspectratio=true]{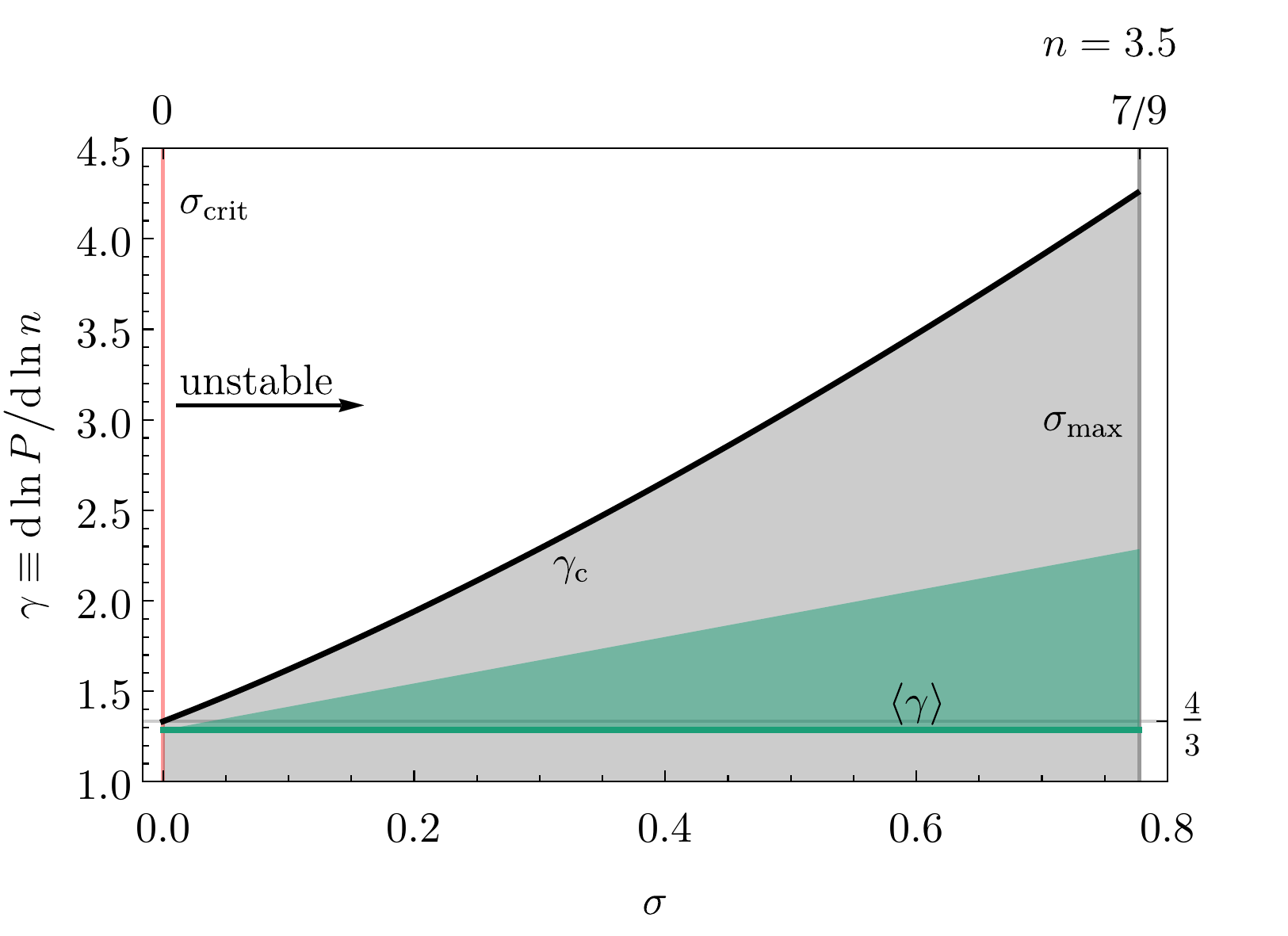}
  \caption{The stability domain as determined by comparison of the effective adiabatic index $\langle\gamma\rangle$ (green line) and the critical $\gamma_\mathrm{c}$ (black light), for polytropic spheres. The red line separates the stable from the unstable (gray) region. The intersection point indicates the maximum relativistic parameter $\sigma_{\mathrm{crit}}$ allowed for stability. The green area indicates the span of $\gamma$ inside the configuration.}
  \label{fig4}
\end{figure}

Using the critical values of the adiabatic index, we computed constraints on the relativistic parameter $\sigma$ in order to construct stable configurations.  We present our results in Fig.~\ref{fig4}, where we plot the ‘effective' adiabatic index, as given by \eqref{effective}, together with the critical adiabatic index $\gamma_\mathrm{c}$. The intersection point determines the critical value of the relativistic parameter $\sigma$ for which the configuration becomes unstable \cite{Nauenberg:1973}. Note how the slope of the curve for $\gamma_\mathrm{c}$ increases as $n$ increases; meanwhile the curve of the effective $\langle \gamma \rangle$ moves downwards until it lies completely below the curve $\gamma_\mathrm{c}$ for $n\geq3$, which indicates that configurations in this regime are unstable for \emph{all} values of $\sigma$. Note also that for $n > 3$, the effective $\langle \gamma \rangle$ approaches to the value $(n+1)/n$ for all $\sigma$.

\section{Energetic considerations and method of critical point to determine radial stability} \label{sec:4}

The energetic approach to the polytrope stability has been exposed in the seminal paper by Tooper \cite{Tooper:1964} on the investigations of polytropic spheres governed fully by general relativistic laws. Bludman \cite{Bludman:1973} refined and extended some of the results of Tooper. In those works, the polytropic spheres were considered for a limited range of the polytropic index, namely $n \leq 3$. Here we present results for radial stability via the critical point method \cite{Shapiro:1983du} for the whole range of the acceptable polytropes, and compare them with our results in Section~\ref{sec:5}.

\subsection{Gravitational and binding energy of polytropes}\label{sec:4.1}

In the relativistic theory the total energy $E$ of certain configuration, which includes the internal energy and the gravitational potential energy, is $Mc^2$ where $M$ corresponds to the gravitational mass producing the gravitational field
\begin{equation}
  E =  Mc^2 = 4\pi c^2\int_{0}^{R}\rho r^2\, \mathrm{d}r\, .
\end{equation}
The proper energy $E_{0}$ is defined as the integral of the energy density over the proper volume, which for a spherical fluid takes the form
\begin{equation}
  E_{0} =  4\pi c^2\int_{0}^{R}\rho e^{\Phi_0} r^2\, \mathrm{d}r\, ,
\end{equation}
where $e^{\Phi_0}$ is given by \eqref{grr}. These two quantities define the gravitational potential energy
\begin{equation}
  E_{\mathrm{p}} = E - E_{0}\, .
\end{equation}
Considering that $e^{\Phi} \geq 1$, we arrive at $E_{0} > E$, therefore the gravitational potential energy is negative. Tooper \cite{Tooper:1964} identifies $E_{\mathrm{p}}$ with the work that must be done \emph{on the system} in order to disperse its constituents to infinity against the gravitational interaction.

\begin{figure}
  \centering
  \includegraphics[width=0.485\textwidth,keepaspectratio=true]{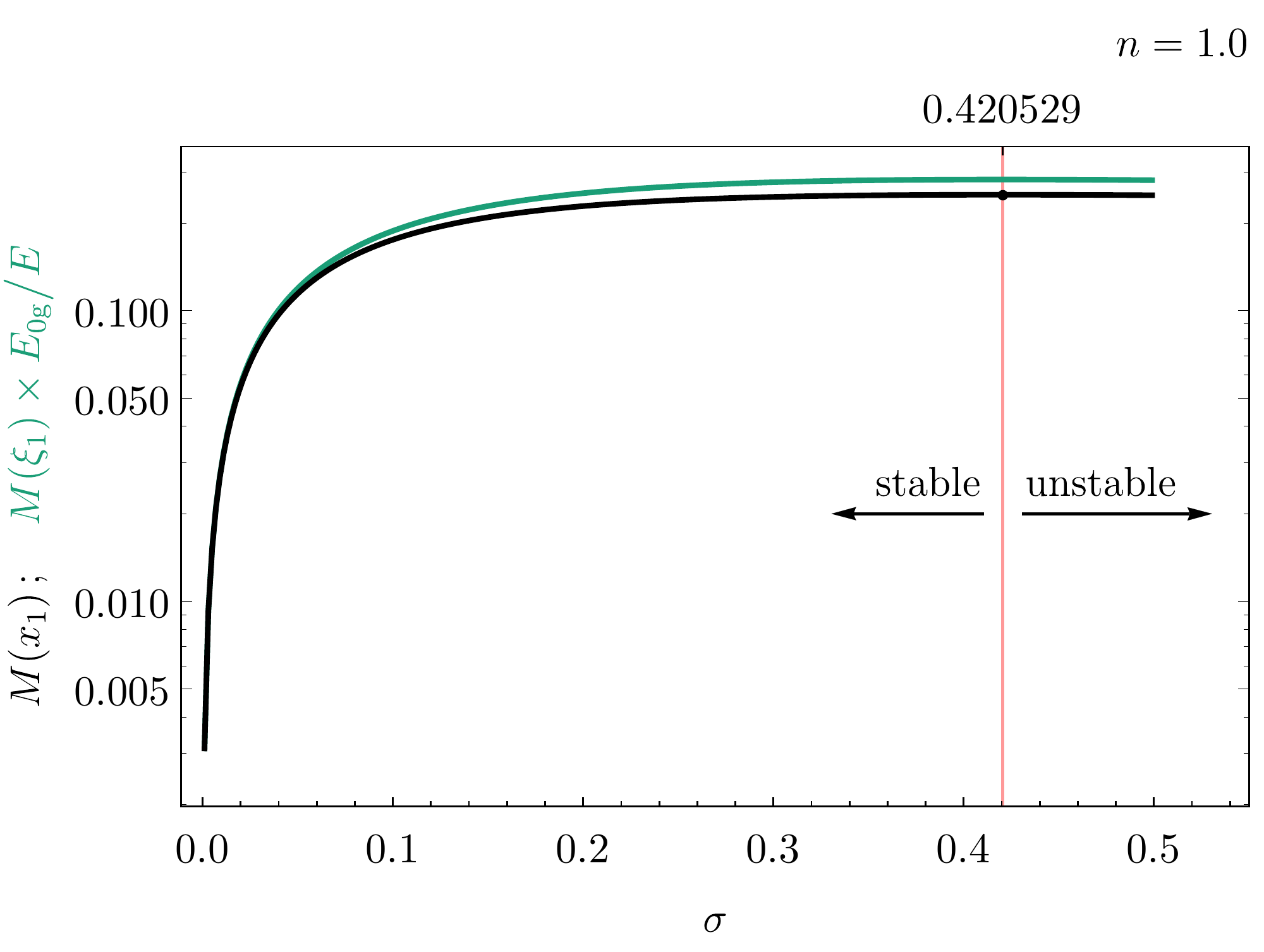}\hfill \includegraphics[width=0.485\textwidth,keepaspectratio=true]{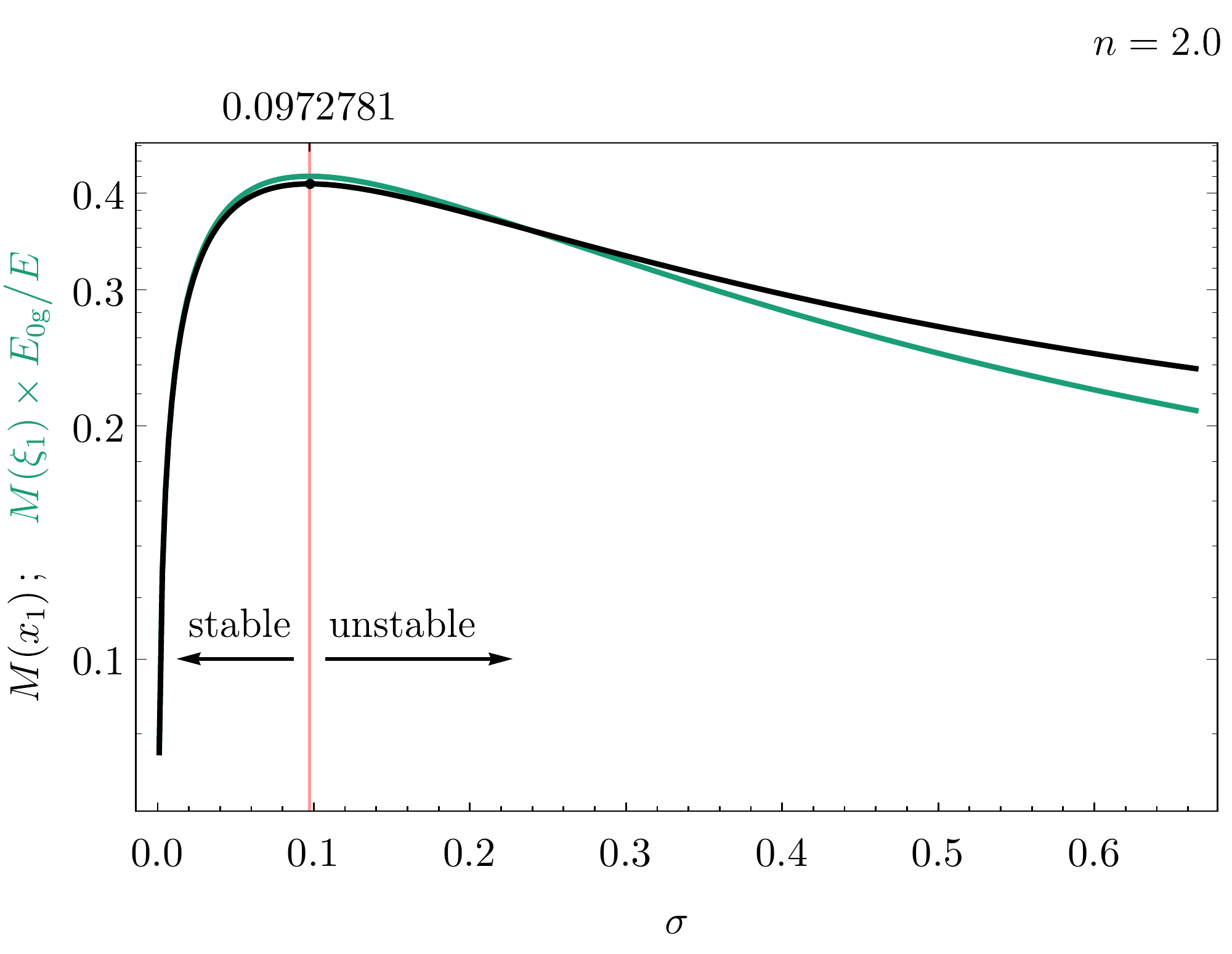}\\
  \includegraphics[width=0.485\textwidth,keepaspectratio=true]{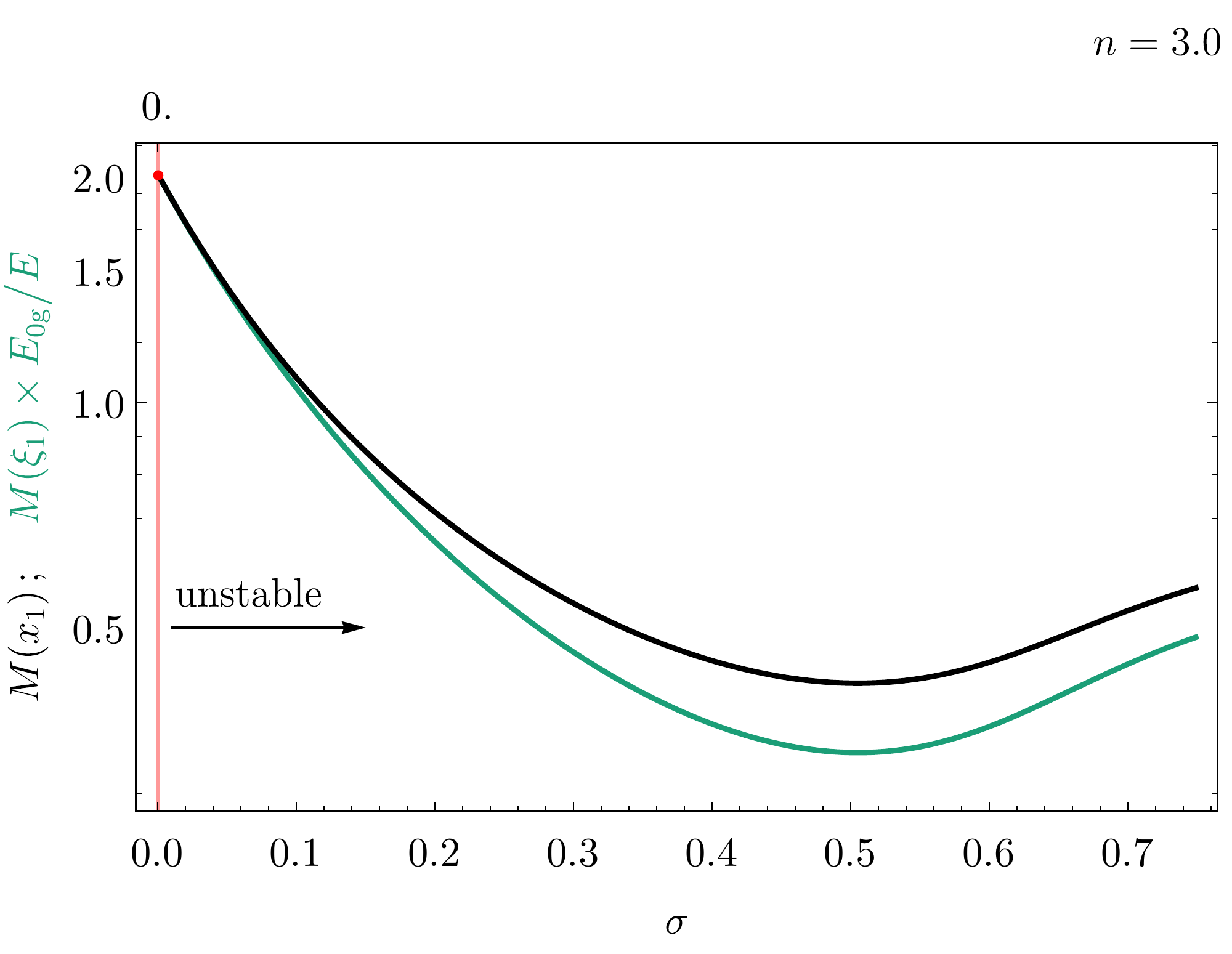}\hfill \includegraphics[width=0.485\textwidth,keepaspectratio=true]{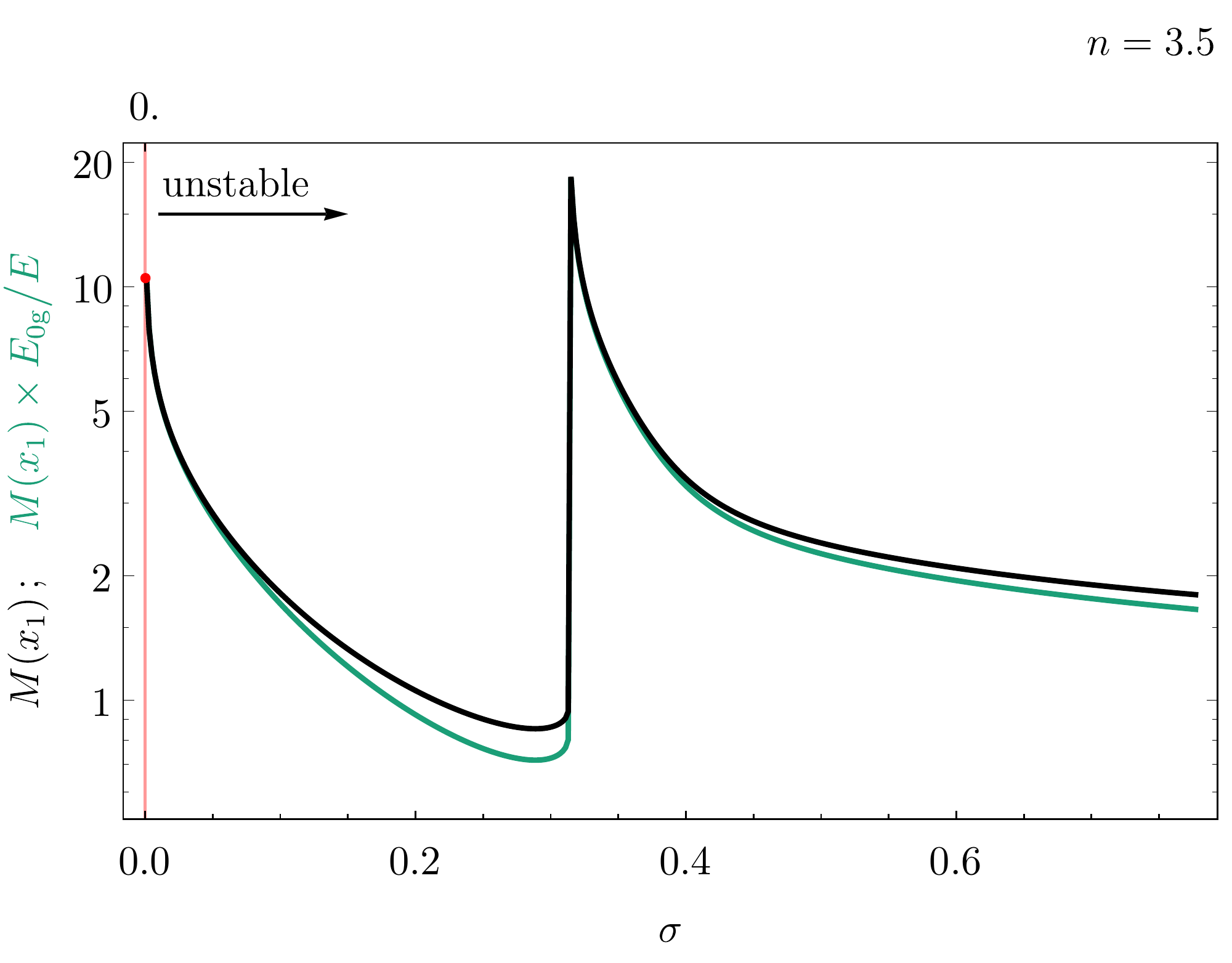}
  \caption{The dimensionless total mass $M$ and rest mass $M_{0g}$, as a function of the relativistic parameter $\sigma$, for polytropic spheres with different values of $n$. We provide the critical value $\sigma_\mathrm{c}$ for stability, as determined by the turning point of the plot.}
  \label{fig5}
\end{figure}

The proper energy and proper mass of a spherical polytrope are defined by
\begin{equation}\label{M0g}
  E_{0\mathrm{g}} =  M_{0g}c^2 = 4\pi c^2\int_{0}^{R}\rho_{g} e^{\Phi_0}\, r^2\,\mathrm{d}r\, ,
\end{equation}
where $M_{0\mathrm{g}}$ indicates the total rest mass of baryons in the configuration, and $\rho_\mathrm{g}c^2$ is the rest energy density of particles. A quantity which will be of importance in our analysis is the ratio
\begin{equation}\label{ratio}
  \frac{E_{0\mathrm{g}}}{E} = \frac{1}{v(x_{1})}\int\limits_{0}^{x_{1}} \frac{\theta^{n}x^2}{[1+\sigma\theta]^{n}[1 - 2\sigma(n + 1) v/x]^{1/2}}\,\mathrm{d}x\, .
\end{equation}
These two quantities define the binding energy of the system, namely, $E_{\mathrm{b}}=E_{0\mathrm{g}} - E$, which corresponds to the difference in energy between an ‘initial' state with zero internal energy where the particles that compose the system are dispersed, and a ‘final' state where the particles are bounded by gravitational interaction.

We follow the standard approach to examine the stability of stars using energy principles (see e.g. Refs. \refcite{Shapiro:1983du,Tooper:1964}). We start by defining the total gravitational mass given by \cite{Tooper:1964}
\begin{equation}
  M =  \frac{1}{4\pi}(n + 1)^{3/2}K^{n/2}(\sigma\, c^2)^{(3 - n)/2} v(\xi_{1})\, ,
\end{equation}
where $K$ and $n$ are the parameters characterizing the polytrope (see~\eqref{polytropic}). Note that we are considering configurations with $K$ and $n$ constant, therefore the total mass $M$ is proportional to $\sigma^{(3 - n)/2} v(\xi_{1})$ and the rest mass is proportional to $\sigma^{(3 - n)/2} v(\xi_{1})(E_{0\mathrm{g}}/E)$.

\begin{figure}
  \centering
  \includegraphics[width=0.485\textwidth,keepaspectratio=true]{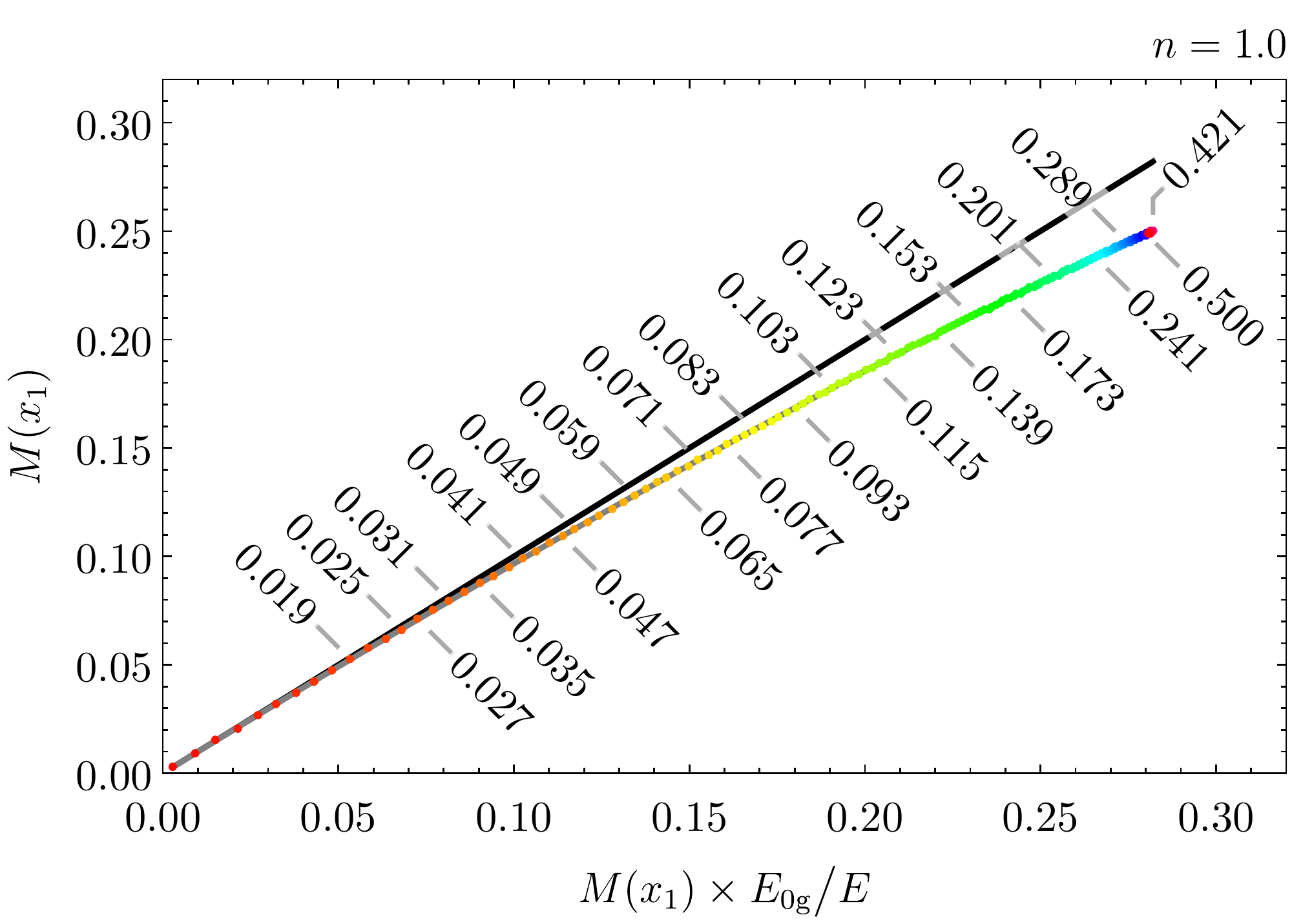}\hfill \includegraphics[width=0.485\textwidth,keepaspectratio=true]{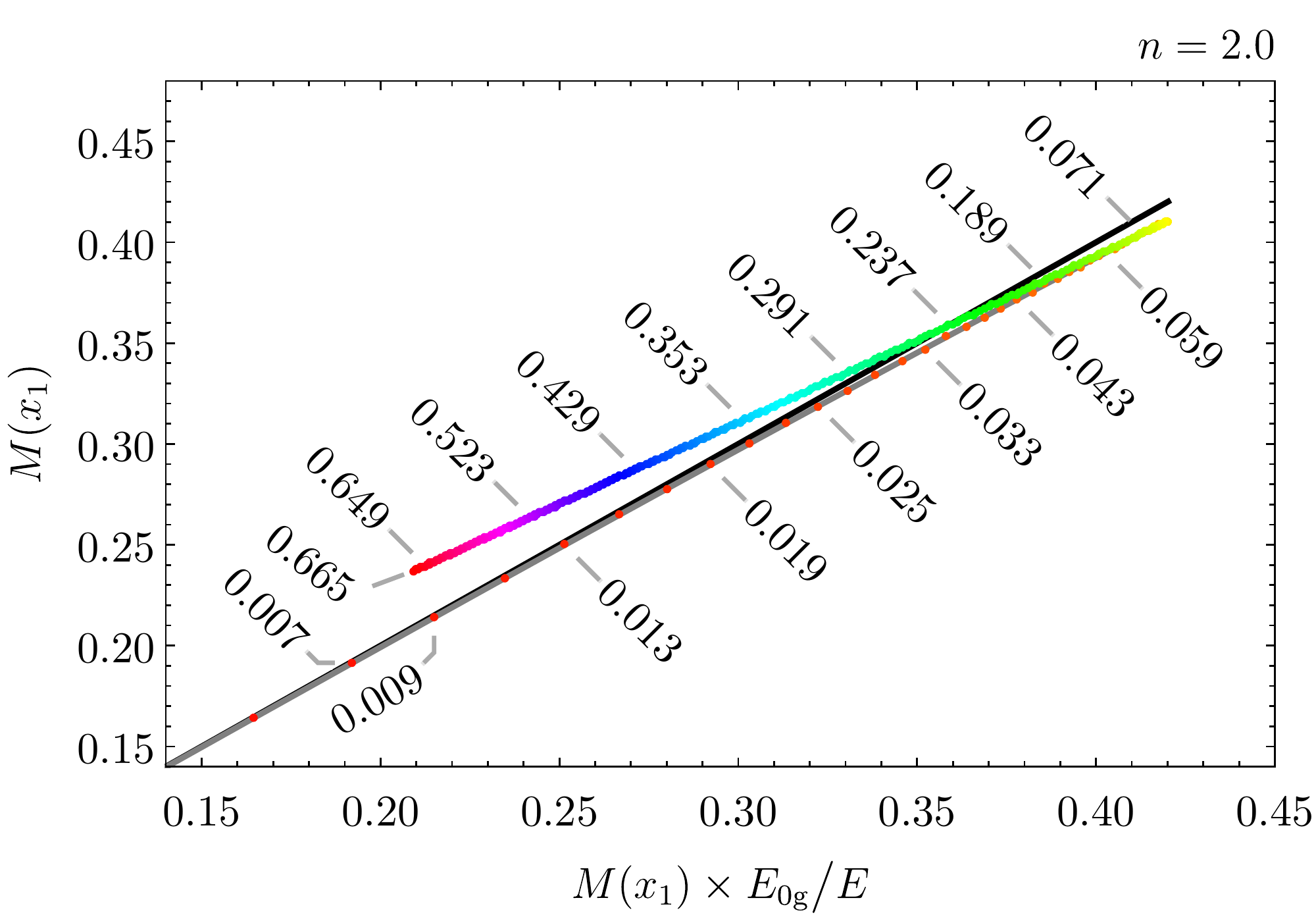}\\
  \includegraphics[width=0.485\textwidth,keepaspectratio=true]{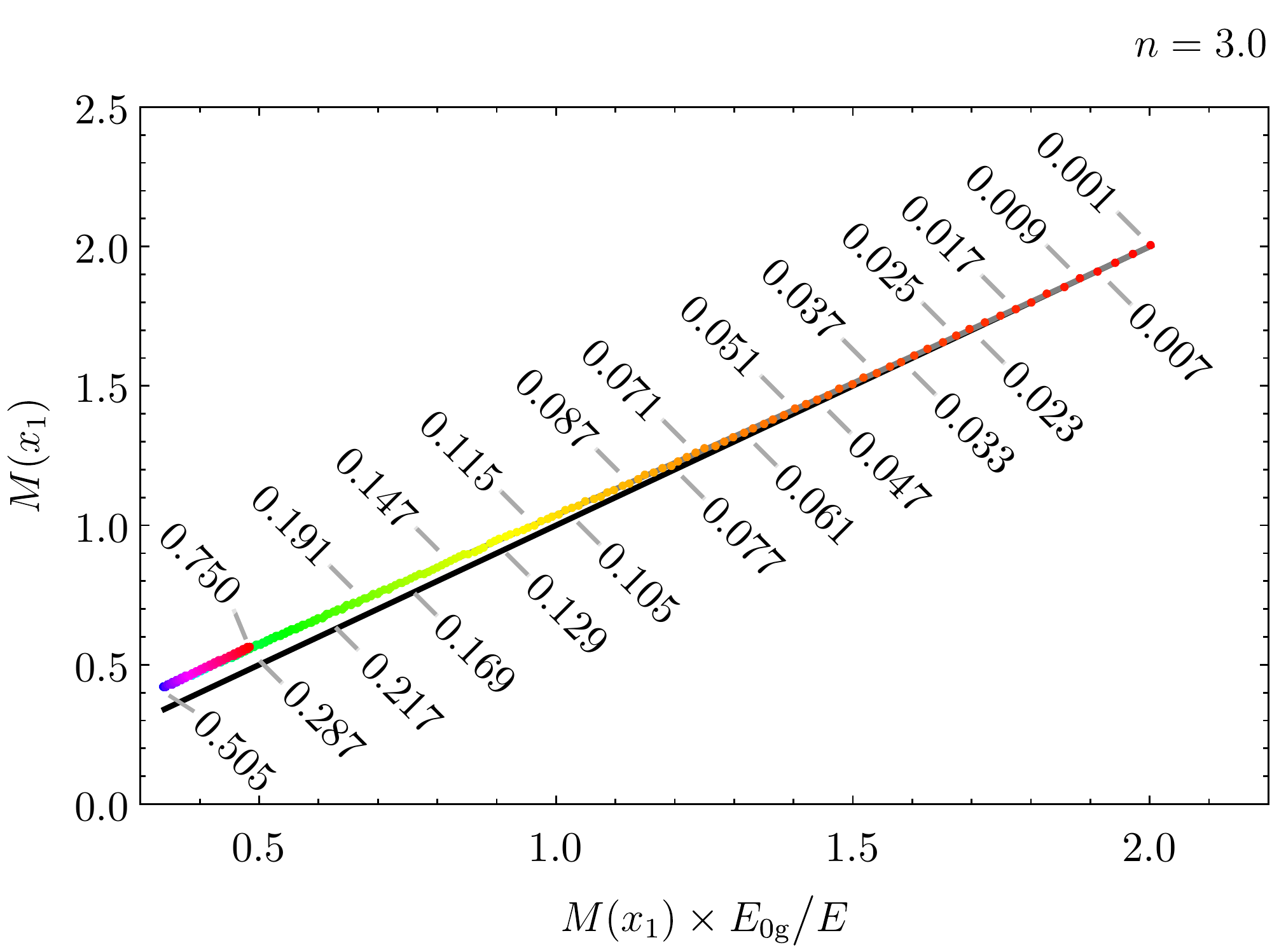}\hfill \includegraphics[width=0.485\textwidth,keepaspectratio=true]{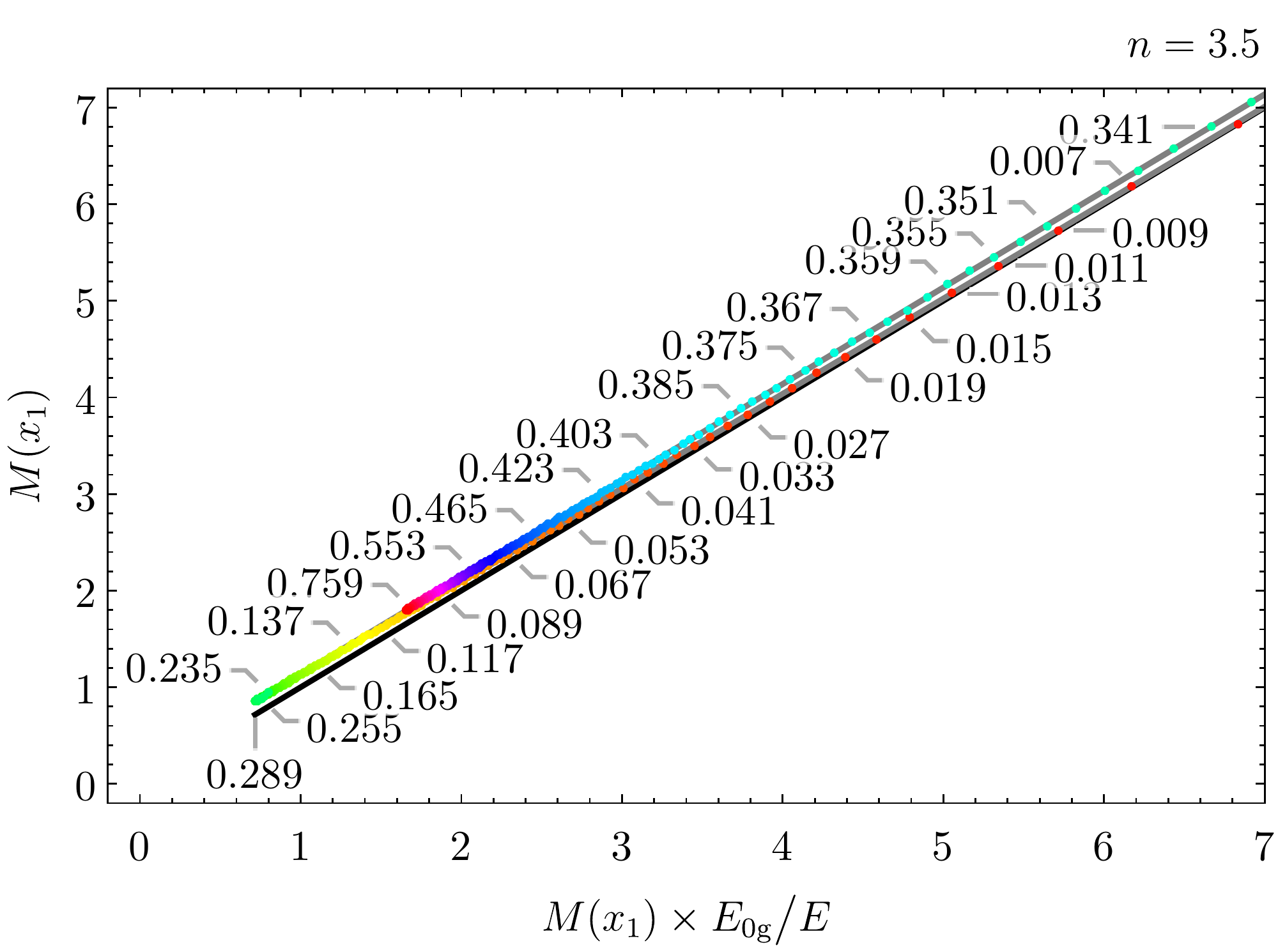}
  \caption{The terminal value of total mass $M$, plotted versus the rest mass $M_{0g} = M(E_{0g}/E)$ for different polytropic spheres. As $\sigma$ increases, there appears a new ‘branch' where two different values of the total mass $M$ correspond to the same values of the rest mass $M_{0g}$.}
  \label{fig6}
\end{figure}

To examine the stability, in Fig.~\ref{fig5} we plot the total gravitational mass $M$ and the rest mass of baryons $M_{0\mathrm{g}}$ (‘preassembly mass'), given by  \eqref{M0g}, against the relativistic parameter $\sigma$. A necessary, but not sufficient, condition for stability is
\begin{equation}
  \left(\frac{\mathrm{d}M_{\mathrm{eq}}}{\mathrm{d}\rho_\mathrm{c}}\right)_{S = \mathrm{const}} > 0\, ,
\end{equation}
where $M_{\mathrm{eq}}$ indicates the total mass at equilibrium. At the critical point, where
\begin{equation}
  \frac{\mathrm{d}M_{\mathrm{eq}}}{\mathrm{d} \rho_\mathrm{c}} = 0\, ,
\end{equation}
there is a change in stability due to the change in the sign of $\omega^2$ (see Section~\ref{sec:3}). Therefore, the critical point where the total mass $M$ demonstrates a maximum indicates the critical value $\sigma_{\mathrm{crit}}$ for stability. We have provided the critical values $\sigma_{\mathrm{crit}}$ for different polytropic spheres with different index $n$ in Fig.~\ref{fig6}.

Notice also in Fig.~\ref{fig5} that a negative binding energy is not a sufficient condition for instability. The binding energy was examined by Zel\nobreak\hspace{-.1em}'dovich and Novikov~\cite{Zeldovich:1971}. For the polytrope $n = 2$, we observe that there is a region with positive binding energy beyond the critical $\sigma$ which is unstable. This example shows some of the limitations of the energetic considerations in the analysis of radial stability.

\begin{figure}
  \centering
  \includegraphics[width=0.9\textwidth,keepaspectratio=true]{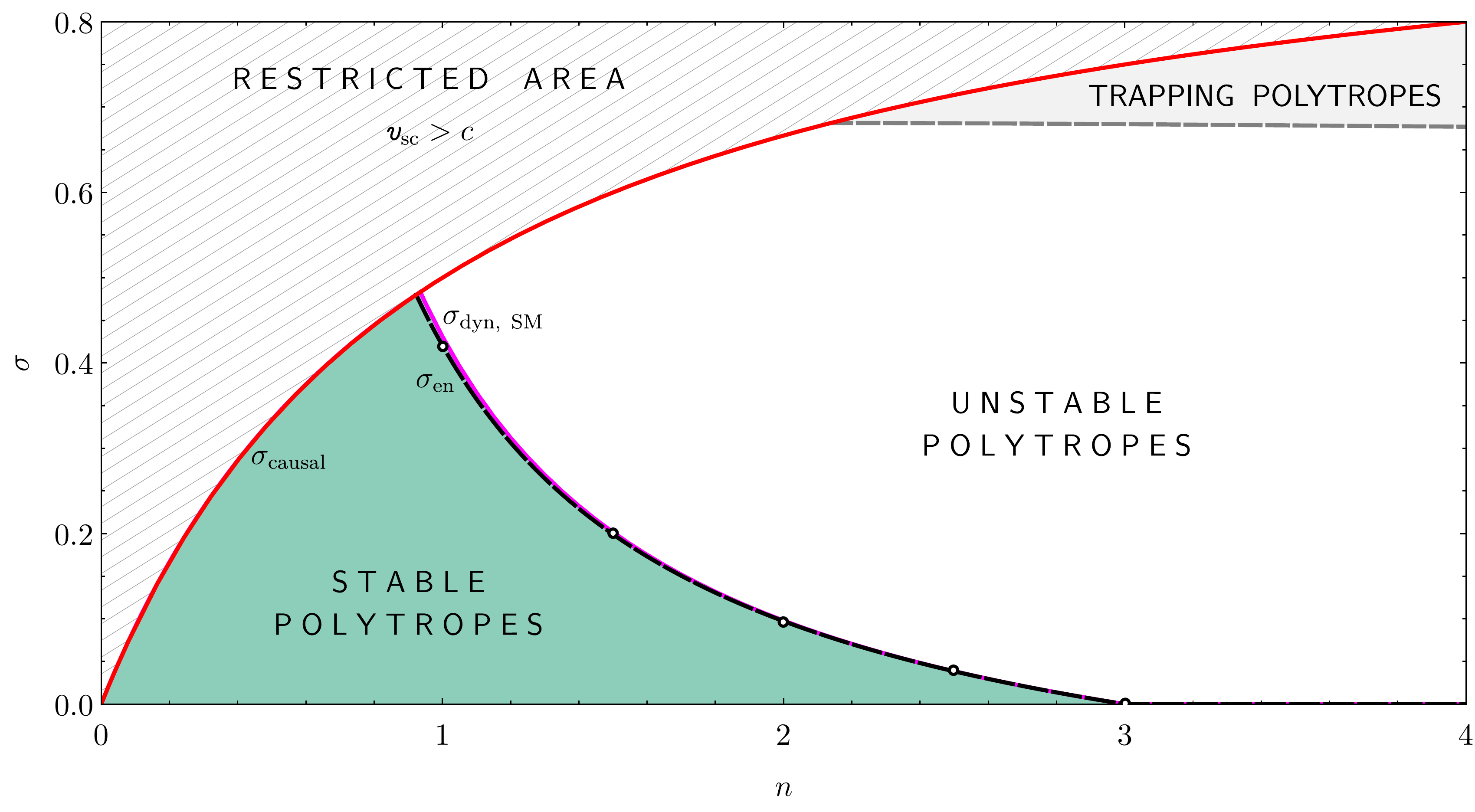}
  \caption{The stability domain for relativistic polytropic spheres as predicted by two different methods: the dynamical approach proposed by Chandrasekhar~\cite{Chandrasekhar:1964zz} computed using the shooting method (magenta line), and energetic considerations (dashed black line). The dots indicate the values obtained by Bludman~\cite{Bludman:1973}. The region of trapping polytropes is also depicted. Configurations in the range $3<n<5$ are unstable.}
  \label{fig7}
\end{figure}

\begin{figure}
  \centering
  \includegraphics[width=0.6\textwidth,keepaspectratio=true]{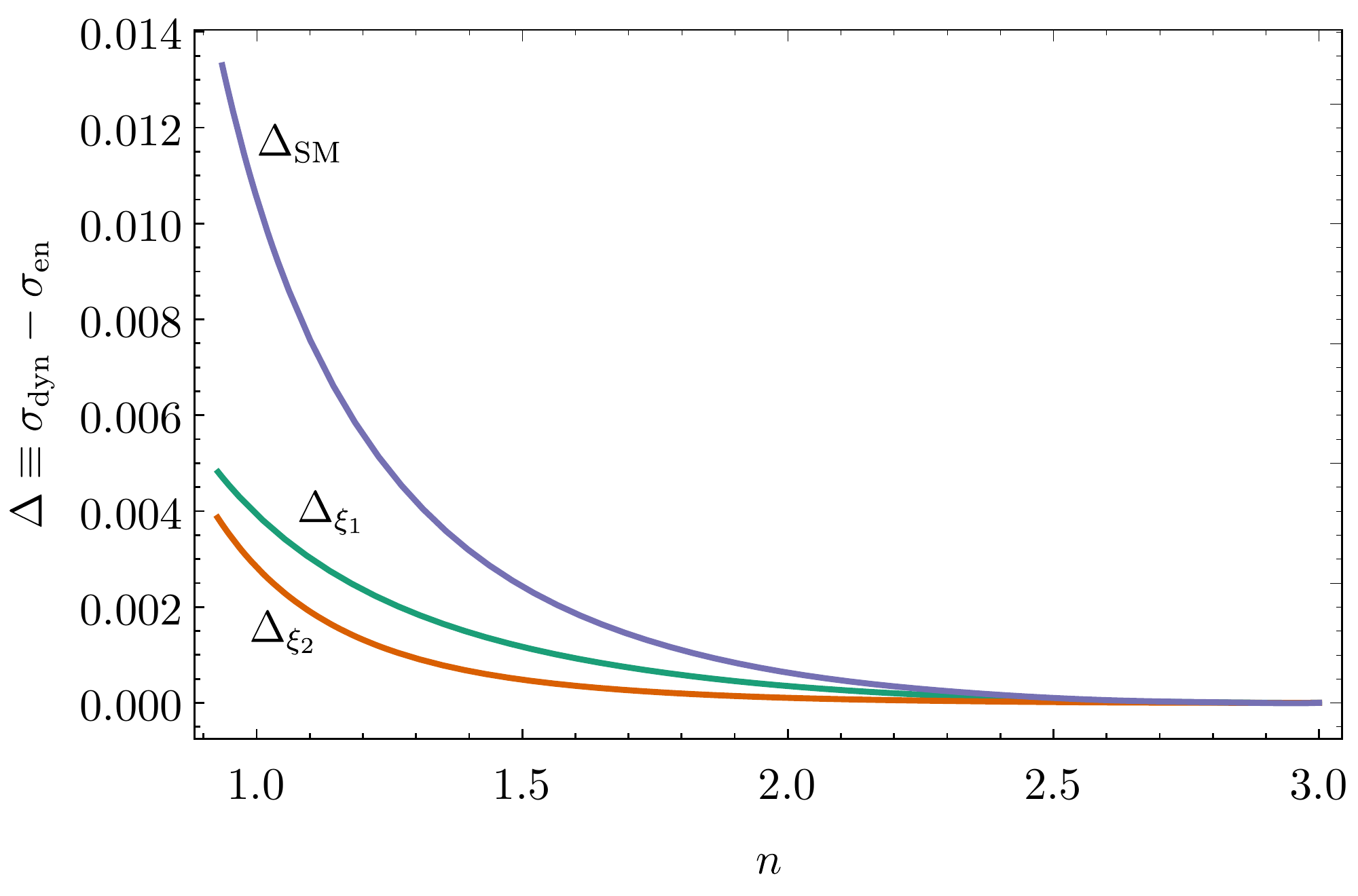}
  \caption{Differences between the maximum allowed parameter $\sigma$ providing stable configurations, for given polytropic index $n$, obtained by dynamical method (all three approaches using trial functions $\xi_1$, $\xi_2$ and the shooting method (SM) are shown) and the results obtained using energetic considerations. Depicted differences are for such $n$ for which allowed $\sigma$ for stability is not given by $\sigma_\mathrm{causal}$.}
  \label{fig8}
\end{figure}

The final result of our analysis gives Figure~\ref{fig7} where we determine stable and unstable region in $(n, \sigma)$ parameter space. In the plot we have included the results provided by the Chandrasekhar method, computed numerically using the shooting method. Results obtained  using the trial functions $\xi_1$ and $\xi_2$ \eqref{Chandratrial} are similar to this one and the differences between results obtained using solving pulsation equation and method using energetic considerations are given in Figure~\ref{fig8}.

Results from all methods are in the good agreement mutually, biggest differences occur for configurations of lower $n$. These results confirm the equivalence between the variational principle derived from the full Einstein equations by Chandrasekhar, and the energy considerations or critical point approach (see Appendix B in Ref. \refcite{Harrison:1965}).  Note that polytropes with $n \geq 3$ are unstable for any $\sigma \geq 0$. The region of trapping polytropes studied in detail in Ref. \refcite{Novotny:2017cep} is fully localised in the unstable region.

\section{Concluding remarks}\label{sec:5}
In order to extend studies of the trapping polytropic spheres~\cite{Stuchlik:2016xiq} whose properties, interesting from the astrophysical point of view, were discussed in a series of works~\cite{Novotny:2017cep,Hod:2018kql,Stuchlik:2017qiz}, in this paper we have studied the radial stability of the polytropic spheres, proposed originally by Tooper, via two different methods, namely, the dynamic method for radial oscillations~\cite{Chandrasekhar:1964zza,Chandrasekhar:1964zz}, and the energetic critical point approach~\cite{Bludman:1973} in the complete range of the polytropic index $0 < n < 5$~\cite{Nilsson:2000zg}. The main conclusion of our work is that the critical value of the relativistic parameter $\sigma$ for the onset of instability is predicted by both methods to be nearly the same. However, we have found, surprisingly, that the predictions of the applied methods have bigger differences for values of $n$ close to $n = 1$. Differences for lower values of $n$ are not manifested because for these configurations the critical value of $\sigma$ is greater than the causality limit $\sigma_{\mathrm{causal}}$. The reason for the obtained differences can not be fully explained only as numerical error.

Despite these differences, we consider that the shooting method provides a very convenient numerical technique to study the stability of relativistic polytropes, considering that it does not require the assumption of certain trial eigenfunction as generally done in the literature. We believe that the shooting method provides a better way to compute the critical values of the adiabatic index, without the vagueness in choosing different trial functions.

Deviations of the estimates of $\sigma_\mathrm{crit}$ for instability, as predicted by Chandrasekhar's dynamical approach and energetic considerations, are not exceeding few percents. On the same level are the differences between the values of $\gamma_\mathrm{c}$ given by using different trial functions and shooting method in the dynamical radial pulsation equation. Moreover, both methods predict that polytropes $n \geq 3$ are unstable for all values of $\sigma$. We also found that the trapping polytropic spheres are localized in the unstable region. This finding may be further refined if other processes are considered, which could have a stabilizing effect, e.g. thermal fluxes, rotation, etc.

Our result concerning the instability of the trapping polytropes against radial pulsations is in agreement with certain indications given by the instability of the trapping polytropes against gravitational perturbations presented in Ref. \refcite{Stuchlik:2017qiz}. Extension of the stability studies to polytropes in spacetimes with a non-zero cosmological constant $\Lambda > 0$~\cite{Stuchlik:2016xiq}, indicates modifications related to the fact that the degeneracy of the polytropes with $\Lambda = 0$ (discussed, e.g. in Ref. \refcite{Bludman:1973}) is broken by the presence of $\Lambda \neq 0$ and the polytrope structure equations directly depend on central density $\rho_\mathrm{c}$~\cite{Stuchlik:2016xiq}. This extension will be considered in a subsequent paper.

\section*{Acknowledgments}
  The authors acknowledge the institutional support of the Faculty of Philosophy and Science of the Silesian University in Opava, and its Research Centre for Theoretical Physics and Astrophysics. C.\,P.\ expresses deep appreciation to John C. Miller for invaluable discussions.

\bibliographystyle{ws-ijmpd}
\bibliography{main}

\end{document}